\documentclass[12pt]{article}

\usepackage[utf8]{inputenc}
\usepackage[T1]{fontenc}
\usepackage{amsmath,amssymb}
\usepackage{booktabs}
\usepackage{tabularray}
\UseTblrLibrary{booktabs}
\usepackage{siunitx}
\usepackage{graphicx}
\usepackage[margin=1in]{geometry}
\usepackage{setspace}
\usepackage{natbib}
\usepackage{url}
\usepackage{hyperref}
\usepackage{placeins}
\usepackage{multirow}
\usepackage{pdflscape} % landscape orientation of table pages
\usepackage{threeparttable}
\usepackage{graphicx}
\usepackage{xcolor}
\usepackage{setspace}
\usepackage{natbib}
\usepackage{titling}
\usepackage{hyperref}
\usepackage{enumitem}
\usepackage{caption}
\usepackage{caption}
\usepackage{pdfpages}
\usepackage{subcaption}
\usepackage{tabularx}
\usepackage{comment}

\usepackage[nameinlink]{cleveref}
\newcommand{\sym}[1]{\rlap{$^{#1}$}}

\hypersetup{
  colorlinks=true,
  linkcolor=blue,
  citecolor=blue,
  urlcolor=blue
}

\usepackage{titlesec}
\titleformat{\subsection}{\normalfont\bfseries}{\thesubsection}{1em}{}

\newcommand{\SSD}{Menu SD}
\newcommand{\BSD}{Binary SD}

\newtheorem{result}{Result}
\newtheorem{hypothesis}{Hypothesis}

\graphicspath{{Output/Figures/}}

\usepackage{titling}
\title{\textbf{Menu Design Meets Market Design: \\Strategic Simplicity vs. Choice Complexity}\thanks{The study was pre-registered at OSF (\url{https://osf.io/r97xh}) and received ethics approval from the German Association for Experimental Economic Research (\url{https://gfew.de/ethik/XRh5gEJX}). Financial support by Deutsche Forschungsgemeinschaft through CRC TRR 190 (Hild, Meisner) and by the National Natural Science Foundation of China (Khanna, Grant W2433170) is gratefully acknowledged. We thank audiences in Berlin and Munich, and, in particular, Christian Basteck, Rustam Hakimov, Siqi Pan, and Clemens Scholz for useful comments and suggestions.}}

\author{
Gian Caspari\thanks{Department of Market Design, ZEW - Leibniz Centre for European Economic Research. Email: \texttt{gian.caspari@zew.de}.}
\and
Paul Ergün Hild\thanks{Humboldt-Universit\"at zu Berlin, Berlin School of Economics. Email: \texttt{paul.hild@hu-berlin.de}.}
\and
Michael Hilweg-Waldeck\thanks{Julius-Maximilians-Universit\"at W\"urzburg. Email: \texttt{michael.hilweg-waldeck@uni-wuerzburg.de}.}
\and
Manshu Khanna\thanks{Peking University HSBC Business School. Email: \texttt{manshu@phbs.pku.edu.cn}.}
\and
Vincent Meisner\thanks{Humboldt-Universit\"at zu Berlin, Berlin School of Economics. Email: \texttt{vincent.meisner@hu-berlin.de}.}
}

\date{\footnotesize\today}

\begin{document}
\maketitle

\vspace{-3em}

\begin{abstract} \noindent
Strategic simplicity removes the incentive to misreport preferences, but it does not ensure that participants can identify what they prefer. We study this distinction in a laboratory experiment on serial dictatorships, where optimal play only requires choosing one's favorite bundle of real-effort tasks. Participants make fifteen incentivized choices among bundles that vary in how many tasks they combine. They choose either from a large menu or through a series of binary choices. Having made all choices, they unexpectedly reconsider each assignment against a previously rejected alternative. We find that roughly one in six choices is reversed and almost one in five involves an objectively dominated bundle. Simplifying the choice architecture does not help. Binary decomposition even raises reversal rates among low-ability participants and further inflates their confidence about their choices.
\end{abstract}


% ============================================================================

\newpage
\doublespacing

\section{Introduction}
\label{sec:intro}

Centralized allocation mechanisms ask participants to report what they want. Whether they can do so reliably, and how the answer depends on the complexity of what they are asked, is a central concern of market design. The canonical response has been to simplify incentives: under a strategy-proof mechanism, truthful reporting is optimal regardless of what others do, and behavioral refinements such as (strong) obvious strategy-proofness \citep{Li2017,PyciaTroyan2023} make this optimality straightforward to recognize. Accordingly, when participants deviate from optimal play, the standard diagnosis is incentive complexity: they fail to see that honesty is  best. In contrast, we consider a setting in which truthful play is obvious, allowing us to focus on the complexity of the choice process itself: whether individuals can reliably identify their preferred options.

In this vein, our paper cautions against hastily attributing puzzling behavior to strategic mistakes. For instance,  in New York City's strategy-proof high-school match, about 7 percent of families file a reapplication after learning their assignment, and more than 70 percent of these reapplicants \emph{reverse} rankings they themselves submitted months earlier, requesting a school they had ranked below, or omitted from, their original list \citep{Narita2018}. In Hungary's strategically simple college admissions system, 11 percent of applicants forgo the free option of applying for a tuition waiver worth thousands of dollars---a revealed dominated choice \citep{ShorrerSovago2023}. Both patterns have been attributed to misunderstood incentives. However, our paper shows that reconsiderations and dominated choices also arise in strategically trivial settings.

How the design of matching mechanisms affects the complexity of the choice process remains understudied. Menu design is an emerging agenda that holds the allocation rule fixed while varying the choice architecture \citep{CaspariKhanna2024, BoHakimov2024}. In this paper, we use a controlled experiment to study a prominent and appealingly simple remedy: binary decomposition replaces a single choice from a large menu with a sequence of binary comparisons to improve choice consistency.

In our experiment, subjects are assigned real-effort tasks in a five-person sequential serial dictatorship, i.e., they make choices in a specified order and are assigned the task they select among the remaining options upon their turn. We isolate the choice architecture through two treatment conditions: Subjects in condition \textit{\SSD{}} select from the entire menu of remaining options, whereas the subjects in condition \textit{\BSD{}} face a series of binary choices. Each participant makes fifteen choices across three complexity blocks. In each block, the five participants cycle through each priority position, and, by construction, face smaller menus in later positions. Across blocks, the complexity of the bundles increases: bundles contain one task in the first block, two tasks in the second, and three tasks in the third block. That is, we compare choice architecture across subjects, but choice complexity (bundle and menu size) within subject.

Standard induced-value designs assign each participant a monetary payoff for every alternative. However, such a design is not suitable for our research question as it assumes away frictions of interest by construction: it directly builds a complete and transitive preference into the experiment and makes alternatives unnaturally easy to compare. Therefore, we have real-effort tasks that are experienced, and choices are incentivized not because of induced values, but because a successful completion of an assignment yields a bonus payment. Preferences over such bundles are real and our data suggest that participants report them inconsistently.

Strategically, sequential serial dictatorships are straightforward: optimal play requires only selecting the best remaining bundle without reasoning about other players' moves. Indeed, both our mechanisms are obviously strategy-proof. \SSD{} is even strongly so, whereas \BSD\ is not (see the discussion in Section \ref{sec:discussion}). Under complete path-independent preferences the two formats implement identical choices from identical feasible sets, so any behavioral difference identifies the effect of the choice architecture.

The central design feature is an unexpected reconsideration stage: after all fifteen choices, each assignment is paired against a previously rejected alternative, so that an incentivized reversal directly measures choice inconsistency. We study these \textit{choice flips} and also record objective \textit{dominance violations} as a model-free measure of mistakes. Section~\ref{sec:design} gives the details.

Before studying our primary hypotheses, we document that choices are inconsistent. Participants flip 17.0 percent of choices under \SSD{} and 18.4 percent under \BSD{}, and roughly 19 percent of initial choices involve an objectively dominated bundle. These magnitudes are comparable to the field evidence from New York and Hungary, and they arise in a setting without strategic motives. For comparison, \cite[Table 4]{Li2017} finds rates of obviously dominated choices less than half as large in his induced-value experiment. In contrast to the school-choice setting in \cite{Narita2018}, our reversals occur within sessions such that they cannot reflect changing circumstances (such as moving districts), new information about schools, or true preference changes over months.

Our central result contradicts our predictions. Binary decomposition does not help on average, and it harms those it is meant to help. Flip rates, dominance violations, and payoffs are on average indistinguishable across the two mechanisms after multiple-testing corrections. However, this average hides effects of opposite signs for low- and high-ability participants, defined as those who perform below and above the median in the tasks, respectively. Among low-ability participants, \BSD{} raises the flip rate from 17.2 to 23.5 percent, an increase of more than a third over their \SSD{} baseline. In contrast, high-ability participants benefit modestly, and only when bundle complexity is low. Decomposing a choice into simpler steps is therefore not a general-purpose remedy: its effect depends on who is choosing.

However, binary decomposition creates an illusion of understanding because it raises the confidence of participants whose choices it makes less consistent. \BSD{} raises reported confidence overall, but the effect is driven mostly by low-ability participants: the same group whose choices become {less} consistent reports feeling substantially {more} sure of them. This pattern is not explained by time on task, and
residual choice inconsistency under \BSD{} persists after controlling for reported confidence.

Next, we identify {which} complexity margin matters. Because the design varies bundle complexity (one to three tasks) and menu complexity (seven down to three alternatives) independently, we can separate two notions that are usually conflated. Choice inconsistency varies with the number of alternatives in the menu, but not with the number of tasks in a bundle. One-task rounds exhibit the highest flip rates. The menu-size effect is limited to menus with few alternatives: reversals are least frequent in the three-bundle menu and do not rise further from four to seven bundles.

\paragraph{Related literature:}
This paper connects insights from decision theory to the experimental literature on market design (see \cite{HakimovKubler2021} for an overview). Starting with \citet{chen2006school}, the early market-design experiments focused on a comparison of different allocation rules to validate theoretical insights. We contribute to the more recent literature on behavioral market design, which focuses on deviations from classical predictions (see \cite{ReesJonesShorrer2023} for an overview). For example, there is ample evidence of suboptimal reporting in strategy-proof assignment mechanisms \citep{hassidim2017mechanism}, and this behavior is also documented in real-life participants of such mechanisms, e.g., of the National Residency Matching Program \citep{ReesJones2018,rees2018experimental}. Moreover, \citet{ShorrerSovago2023} record dominated choices in a strategically simple admissions environment, and \citet{Narita2018} shows that families in NYC's strategy-proof school assignment reverse their own submitted rankings upon reapplication. The behavioral market-design literature has branched off into several explanations for such observations, and we relate our paper below.

First, players with standard preferences may misunderstand the mechanism or struggle to see the optimal strategy due to cognitive limitations. \citet{basteck2018cognitive,basteck2023aiding} show experimentally that low-ability participants are disproportionately mismatched under immediate acceptance, and that strategy-proof mechanisms ``level the playing field'' \citep{pathak2008leveling}. \citet{klijn2019static} and \citet{bo2020iterative} show that sequentializing the allocation process (via iterative DA) improves truthful play. To formalize strategic simplicity, \citet{Li2017} introduces obvious strategy-proofness, and \citet{PyciaTroyan2023} strengthen this concept. Papers of this literature focus on the strategic structure of mechanisms. Since we find choice reversals in both of our obviously strategy-proof mechanisms, an independent dimension of simplicity seems to be at work: the cognitive reliability of the choice process a mechanism induces.

Second, players might understand the mechanism and have standard preferences, but these preferences may not be known or formed yet. \cite{chen2021information} study students’ incentives to acquire information about their own and others’ preferences. \cite{hakimov2023costly} investigate the interplay of these incentives with provided information on admission chances. Closer to our work, \cite{GrenetHeKubler2022} explain the observation that college program choices in Germany depend on the order in which offers are observed using a model in which information acquisition incentives depend on the availability of offers. In contrast, the subjects in our experiment do not have admission uncertainty, and they reverse choices without learning new information.

Third, although players may perfectly understand the assignment mechanism and know what they want, they might maximize an expected utility function whose optimum depends on their beliefs. For example, they might be expectation-based loss averse \citep{DreyfussHeffetzRabin2022,MeisnerWangenheim2023}, have ranking-dependent preferences \citep{meisner2023report,kloosterman2025rankings}, or hold overconfident beliefs about their own priorities \citep{pan2019instability,meisner2026confidence}. Such biases or preferences would not affect choice behavior in our experimental setting where participants choose under certainty. In contrast, \cite{CaspariKhanna2024} assume that the choice itself, independent of beliefs and the allocation mechanism, can be non-standard, and our paper follows this line of thought.

To the best of our knowledge, \citet{CaspariKhanna2024} are the first to study menu design with a focus on choice behavior. In contrast, the focus of pick-an-object implementations \citep{BoHakimov2024} is on strategic considerations. The emerging \emph{menu design} agenda holds the allocation rule fixed and varies the choice architecture through which preferences are expressed. \citet{BudishKessler2022} ask whether participants can report preferences accurately enough at all.

Finally, our reconsideration stage relates to evidence that repeated choices from similar or identical sets can be inconsistent \citep{AgranovOrtoleva2017}, and to work on choice overload, conflict, and deferral \citep{TverskyShafir1992, Gerasimou2018}. In contrast to standard individual-choice experiments, our setting embeds individual choices in an allocation mechanism: Under serial dictatorship, each participant's choice affects the options available to lower-priority participants and can therefore have implications for market-level outcomes.

The rest of the paper proceeds as follows. Section \ref{sec:design} describes the experimental design. Section \ref{sec:hypotheses} presents the behavioral hypotheses. Section \ref{sec:results} reports the results, Section \ref{sec:discussion} discusses candidate explanations, and Section \ref{sec:conclusion} concludes.

% ============================================================================

\section{Experimental Design}
\label{sec:design}

\subsection{Overview}

The experiment has five stages. First, participants learn about and practice four real-effort tasks. Second, they report subjective evaluations of the tasks and their attributes. Third, they sequentially make allocation choices through a serial dictatorship mechanism, whose implementation depends on the treatment condition. Fourth, they reconsider their choices against previously rejected alternatives, one of which is randomly selected for payment. Finally, participants carry out this randomly selected task and obtain a bonus payment in case of successful completion.

We implemented the experiment in 26 sessions with 494 participants. Of these, 241 participants were assigned to treatment condition \SSD$\text{ }$ and 253 to condition \BSD. Each participant makes 15 mechanism choices: five choices at each of three bundle-complexity levels. With 494 participants and 15 choices per participant, the main choice-level sample contains 7,410 observations. Mechanisms are allocated within sessions: participants are numbered in their order of arrival, and a random draw at the start of each session determines which mechanism is assigned to the first half of arrivals and which to the second half (the second half receives one additional participant when a session has an odd number of participants). Pre-treatment covariates (practice earnings, variety preference, gender) are balanced across mechanisms, with one exception: participants assigned to \BSD{} score $0.25$ standard-deviation units higher on the learning-stage ability index ($p<0.001$).\footnote{The imbalance is concentrated in two sessions whose anomalous learning-stage records all fall in the \SSD{} arm; we retain these participants throughout. Because the imbalance makes the \BSD{} arm more able on average, it works against rather than for the harm to low-ability participants reported in Section~\ref{subsec:results-mechanism-heterogeneity}, and Appendix~\ref{app:robustness} shows that this result is robust to how the imbalance is handled.}

The experiment was conducted at the University of Vienna's laboratory between November 2025 and March 2026. The study was pre-registered and received ethics approval from the German Association for Experimental Economic Research.\footnote{Pre-registration: \url{https://osf.io/r97xh}; Ethical approval: \url{https://gfew.de/ethik/XRh5gEJX}. The experiment was coded in oTree \citep{chen2016otree}. Full screenshots of the experiment can be found in the online appendix.}

\subsection{Learning stage}

Participants first receive instructions about the four tasks described in Subsection \ref{sec:tasks}. Next, they perform each task twice under incentives.\footnote{One round of one task, chosen uniformly at random across the eight learning rounds, determines the practice-stage payment, paid at a task-specific per-correct-answer rate of 15 cents for the math-memory and quiz tasks and 40 cents for the spot-the-difference and emotion-recognition tasks.} The learning stage serves two purposes. First, it familiarizes participants with the tasks before they make allocation choices. Second, it provides individual-level performance measures used to construct ability and expected-payoff measures. After the learning stage, participants are given feedback on their performance across the eight rounds.

\subsection{Qualitative pre-mechanism survey and scores} \label{sec:survey}

Immediately after the practice-stage feedback screen, participants completed an 11-screen qualitative survey eliciting their preferences over the tasks. We elicited preferences over four task-level attributes---\emph{Cognitive Ease}, \emph{Engagement}, \emph{Confidence}, and \emph{Time Efficiency}. Once participants had ranked each task against the four attributes, they ranked the four attributes themselves in a similar fashion. Lastly, we used a 5-point Likert scale to elicit participants' preference between bundles containing different tasks and bundles containing repeats of the same task. The full instrument and procedural details appear in Appendix~\ref{app:qualitative-survey}. 

We use survey responses to construct scores for all tasks, and the exact aggregation rule from survey responses to scores is given in Appendix~\ref{Appendix_Algorithm}. The scores serve as an alternative preference representation as they construct a full ranking of the available tasks. These scores select the alternative offered in the reconsideration stage and define the score-fit and score-deviation outcomes. The variety-preference item enters heterogeneity analyses as a participant-level covariate.

\subsection{Tasks and difficulty} \label{sec:tasks}

In the mechanism stage, the participants choose among bundles composed from the four real-effort tasks they practiced in the first stage:

\begin{enumerate}[leftmargin=*]
  \item \textbf{Math-memory task.} Participants play a memory game on a $4 \times 4$ grid of face-down boxes. A deck consists of sixteen boxes. Each box conceals a simple addition (e.g., $2+1$) and a colored heart (red or black). A pair is two boxes that share both the same numerical sum and the same heart color. Participants click on boxes two at a time to reveal what is behind. Once eight matching pairs are found, a fresh deck appears. Participants may play up to three decks within a two-minute round, and each correct pair scores one point.
  \item \textbf{Spot-the-difference task.} Participants see two near-identical images containing ten minor differences. They obtain one point per difference they click.
  \item \textbf{Quiz task.} Participants answer up to thirty four-option multiple-choice questions on general knowledge drawn from Art, Languages, Geography, Technology, and History. Each question is displayed for a maximum of ten seconds. The answer buttons are disabled for the first four seconds to deter reflexive clicking. A correct answer yields a point.
  \item \textbf{Emotion-recognition task.} Participants are shown ten head-shots of individuals expressing different feelings. The task is to identify the correct emotion from four emotion labels (e.g., “anger,” “sadness,” “joy”). A correct answer yields a point.
\end{enumerate}

An assigned task is performed for two minutes. We vary the difficulty by varying the point threshold required for the task to count as successful. Higher difficulty requires more points on the same task within the same time limit. There are three difficulty levels: 1 (Easy), 2 (Medium), and 3 (Hard). The corresponding score thresholds were set at the 20th, 50th, and 80th percentile of the point distribution observed in previous studies of that task.\footnote{The implemented cutoffs are 8, 14, and 17 of 30 correct answers in the quiz, 12, 16, and 24 of 24 in math memory, 2, 6, and 8 of 10 in spot-the-difference, and 3, 5, and 8 of 10 in emotion recognition.}

Because a fixed bonus is only paid out when the task was successfully completed, a lower-difficulty version of a fixed task objectively dominates its higher-difficulty version. The participants are free to continue for the full two minutes even after the threshold has been reached. The reason behind this design choice is that participants who like the challenge of the higher threshold or who enjoy playing the task as a game can replicate the hard environment.

\subsection{Bundles and complexity}

\begin{figure}[htbp]
    \centering
    % Top Panel: Panel (a)
    \begin{subfigure}{\textwidth}
        \centering
        \includegraphics[width=0.85\textwidth]{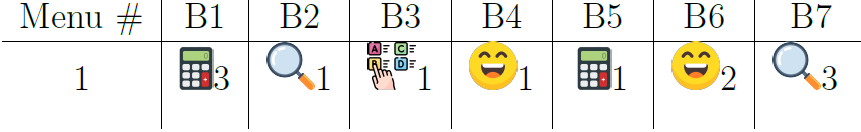} % Replace with your actual file name
        \caption{Low complexity menu example}
        \label{fig:bundle_a}
    \end{subfigure}
    
    \vspace{0.5cm} % Adjust vertical spacing between top and bottom panel if needed

    % Bottom Panel: Panel (b)
    \begin{subfigure}{\textwidth}
        \centering
        \includegraphics[width=0.85\textwidth]{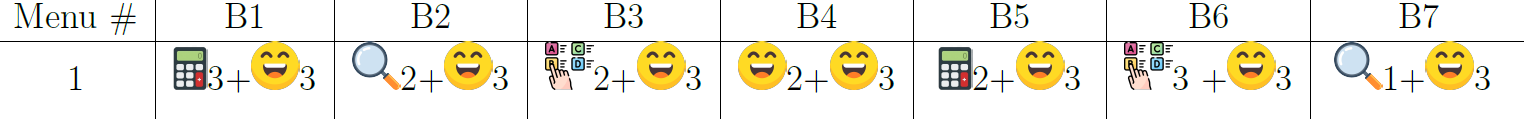} % Replace with your actual file name
        \caption{Medium complexity menu example}
        \label{fig:bundle_b}
    \end{subfigure}

    \caption{Panel (a) depicts an example menu with Low complexity. In these menus, each bundle contains a single task with a particular difficulty level. For instance, the second bundle in the menu is a Spot-the-differences task of easiest difficulty and the 7th bundle is the same task of the highest difficulty. Panel (b) depicts an example menu with Medium complexity. In these menus, each bundle contains two tasks with varying difficulties. For instance, the second bundle contains one Spot-the-differences task of medium difficulty and an  Emotion-recognition task of the highest difficulty.}
    \label{fig:bundle_comparison}

\end{figure}

A bundle is a collection of one, two, or three tasks (Figure~\ref{fig:bundle_comparison}) of various difficulty levels  which are indexed by a subscript 1 (Easy), 2 (Medium), or 3 (Hard). The mechanism-stage bonus is all-or-nothing: a participant who meets the threshold on \emph{every} component task of the assigned bundle earns 20\,EUR, and 0\,EUR otherwise.
We define bundle complexity by the number of tasks in the bundle:
\begin{itemize}[leftmargin=*]
  \item \textbf{Low complexity:} one task per bundle.
  \item \textbf{Medium complexity:} two tasks per bundle.
  \item \textbf{High complexity:} three tasks per bundle.
\end{itemize}

In each round, a group of five participants faces a menu of seven bundles. Since sequential serial dictatorship removes one bundle after each assignment, the participant with priority rank 1 chooses from seven bundles, rank 2 from six, and so on until rank 5 chooses from three bundles.
Therefore, this mechanism features a second, built-in complexity measure: menu size.

Participants are linked into five-person markets at the start of the mechanism stage, and each participant's menus depend on the choices of the same market across all 15 mechanism rounds. 
Rounds are organized into three complexity blocks (Low~$\to$~Medium~$\to$~High). Within each block, the participant cycles through priority ranks 1 to 5, so the menu size shrinks from seven to three within the block. Across the five priority rounds within a block, the composition of the initial seven-bundle menu varies and no two rounds begin from the same set of bundles, so that participants cannot memorize a target across rounds. The initial seven-bundle menus contain strictly dominated decoys, such as B6 in Panel \ref{fig:bundle_a} (dominated by B4) or B1 in Panel \ref{fig:bundle_b} (dominated by B5).

\subsection{Mechanisms}

Participants are assigned to one of two mechanisms.

\paragraph{Menu serial dictatorship.}
In \SSD, participants choose directly from the remaining menu when their priority rank is reached. The highest-priority participant chooses from all seven bundles, the next participant chooses from the six remaining bundles, and so on.

\paragraph{Binary serial dictatorship.}
In \BSD, the participant faces the same remaining menu, but the choice is decomposed into a sequence of binary comparisons. The menu is first split into exclusive pairs that are compared. Next, new binary comparisons are formed from the selected options (and the remainder, if the menu size is odd), and this process continues until the final winner is assigned.

\subsection{Reconsideration stage}

After the 15 mechanism choices, participants enter the reconsideration stage. For each initial assignment, the assigned bundle is paired with an alternative bundle that was available and rejected during the original mechanism choice. Participants choose again between the assigned bundle and the alternative. A reversal is coded as a flip. One reconsidered choice is randomly selected to determine the final bundle, and the participant then performs the tasks in that bundle for a 20 EUR bonus.
Hence, a flip is not mere cheap talk because it changes the participant’s final allocation and expected payoff by changing the probability of obtaining the 20 EUR bonus.

On screen, each reconsideration round is framed neutrally as a binary choice between a ``Left bundle'' and a ``Right bundle,'' without indicating which bundle was the original mechanism-stage selection. In addition, the position of the original selection is randomized between the two positions. Each of the 15 reconsideration choices is equally likely to be selected for payment, so every reconsideration decision is incentivized.
The alternative bundle in each reconsideration round was not random. It was the highest-scoring alternative according to the scoring function constructed from the survey described in Section \ref{sec:survey}.

 This design choice has a direct consequence for the interpretation of flip rates. The reconsideration stage detects deviations between a participant's mechanism-stage choice and the bundle that their elicited preference scores predict as best or second-best.

\subsection{Payment}

Participants receive three components of pay. First, a participation fee of 5\,EUR, fixed and independent of performance. Second, a learning-stage bonus determined by one of the eight learning rounds chosen uniformly at random. The rate is 15 cents per correct answer for the math-memory and quiz tasks and 40 cents per correct answer for the spot-the-difference and emotion-recognition tasks, capped at the task-specific maximum, yielding a learning-stage bonus of up to 4.50\,EUR. Third, an all-or-nothing mechanism-stage bonus of 20\,EUR if the participant meets the threshold on every component task of the randomly selected reconsideration bundle, and zero otherwise. The average participant earned 19 EUR, and 55.7 percent of the participants obtained the bonus payment.

\section{Behavioral Hypotheses}
\label{sec:hypotheses}

Serial dictatorship is the most favorable possible environment for accurate
reporting: participants only have to pick their most preferred bundle independently of others' behavior. Under complete path-independent preferences, truthful choice at each binary comparison selects the same maximal element that the participant would choose directly from the full menu. Moreover, independent of menu size and bundle complexity, this choice should be robust to any reconsideration. Because both \SSD\ and \BSD\ are obviously strategy-proof,\footnote{While \SSD\ is strongly obviously strategy-proof, \BSD{} is only weakly so, but additionally accommodates non-standard choice by satisfying the corresponding incentive criterion in \cite{CaspariKhanna2024}.} any departure from this benchmark suggests a friction in the \emph{choice process} rather than strategic manipulation. 
Our hypotheses below ask if deviations from the benchmark vary with the assigned mechanism and choice complexity.

Let $C_{ir}, R_{ir}, S_{ir}$ be the choice in the mechanism stage, the  choice in the reconsideration stage, and the highest-scoring bundle of participant $i$ in round $r,$ respectively.
Moreover, let $\mathcal D_{ir}$ be the set of dominated task bundles that participant $i$ faces in round $r.$
We measure failures with three round-level indicators. The central outcome is a
flip, 
\[ F_{ir}=1 \quad 
\mbox{ if } C_{ir} \neq R_{ir}, \quad\mbox { and } F_{ir}=0 \quad 
\mbox{otherwise. }\]
Since one reconsideration decision is paid, a flip is an
incentivized change in the final allocation. The alternative offered is the
participant's own elicited second-best bundle (Appendix~A.2), so $F_{ir}$ measures
whether a choice survives a second look against the most tempting feasible
competitor. A dominance violation occurs when the chosen bundle is
objectively dominated by another available one consisting of identical tasks that are all weakly easier and at least one is strictly easier,
\[ D_{ir}=1 \quad 
\mbox{ if } C_{ir} \in \mathcal D_{ir}, \quad\mbox { and } D_{ir}=0 \quad 
\mbox{otherwise. }\]
This is the most direct, model-free notion
of a mistake as it only reduces the chance to receive the bonus without impacting other aspects of the task. 
A score-fit deviation occurs when the choice is not
the highest-scoring available bundle under $i$'s elicited preference scores, 
\[ M_{ir}=1 \quad 
\mbox{ if } C_{ir} \neq S_{ir}, \quad\mbox { and } M_{ir}=0 \quad 
\mbox{otherwise. }\]
We also study expected profit,\footnote{Expected profit is the mean, over a participant's fifteen mechanism-stage choices, of the expected mechanism-stage bonus of the bundle chosen in that round, before reconsideration: $20\,\text{EUR}\times\prod_{t\in b}\big[1-\Phi\big((c_{tl}-\mu_{it})/\sigma_t\big)\big]$, where the product runs over the component tasks $t$ of bundle $b$, $c_{tl}$ is the threshold of task $t$ at difficulty level $l$, $\mu_{it}$ is participant $i$'s mean score over the two learning-stage plays of task $t$, and $\sigma_t$ is the standard deviation of learning-stage scores on task $t$ pooled across participants and plays. It is thus a participant-specific, pre-assignment measure of how much the chosen bundles are worth to that participant, in expectation.}
realized payoff, score fit of the final bundle, and self-reported confidence
and happiness on seven-point scales. Heterogeneity is organized by an
ability index $A_i=\tfrac{1}{8}\sum_{t=1}^{4}\sum_{n=1}^{2} z_{int}$, where $z_{int}$ is participant $i$'s learning-stage score in play $n$ of task $t$, standardized across
participants.\footnote{Standardizing each round before averaging is necessary. The four tasks are scored on very different scales (learning-stage standard deviations of roughly $6.6$ for math-memory, $3.9$ for the quiz, and $1.6$ for both spot-the-difference and emotion recognition), so an unstandardized average of the raw scores would be roughly two-thirds math-memory and a quarter quiz by variance and would not measure average performance across the four tasks. Both incentivized plays of each task enter the index. The four task scores share a common component: a participant below the median on one task is below the median on each of the other three with probability $0.54$ to $0.65$, against $0.50$ under independence (Appendix Table~\ref{tab:transition}).}

Let $P(X)$ denote the empirical frequency of $X$.
We measure choice quality by (the absence of) flips in the reconsideration stage and dominated choices.
The underlying assumption behind our primary hypotheses is that these objects of interest exist in the first place.

\addtocounter{hypothesis}{-1}
\begin{hypothesis}[Choice inconsistency]\label{hyp:inconsistency}
In both serial dictatorships,
choices are inconsistent,
$$P(F_{ir}=1)>0 \quad \mbox{ and } P(D_{ir}=1)>0.$$
\end{hypothesis}

Our study is motivated by \citet{CaspariKhanna2024} who show that pairwise comparisons
require less choice sophistication for pairwise stability than choices from larger menus, which suggests \emph{simplification}: Binary SD should reduce flips and dominance violations, especially in larger and more
complex problems. 
However, a sequence of simple comparisons need not produce a better final choice. Participants may anchor on the incumbent or neglect alternatives they never compare directly as well as become overconfident because each step feels easy. Alternatively, the increase in mouse clicks might come with an increase in clicking mistakes. We subsume all these negative effects under the term \emph{fragmentation}.

\begin{hypothesis}[Simplification versus fragmentation]\label{hyp:mechanism}
Simplification: Binary decomposition improves choice quality 
\[
P^{Binary}(F_{ir}=1) < P^{Menu}(F_{ir}=1) \mbox{ and } P^{Binary}(D_{ir}=1) < P^{Menu}(D_{ir}=1).
\]
Fragmentation: Binary decomposition reduces choice quality so that the inequalities above are reversed.
\end{hypothesis}

Because the two formats are outcome-equivalent under the benchmark, any difference is behavioral. 
The two forces can coexist and offset each other across participants of different ability. 
In fact, we believed that high-ability participants gain or lose little from binary decomposition, as they are able to identify their most-preferred bundle relatively easily.
In contrast, we expected low-ability participants to find selection more difficult.

The second hypothesis separates two aspects usually conflated under ``complexity.'' Bundle
complexity (one, two, or three tasks) raises the {dimensionality} of each
alternative, and menu size (which shrinks from seven to three as priority falls) raises
the {number} of alternatives to choose from. Theories of attribute integration point to the
former, theories of choice overload to the latter, and the design lets them move
independently.

\begin{hypothesis}[Complexity]\label{hyp:complexity}
Flips and dominance violations increase with bundle size and menu size, 
$$
P(F_{ir}=1) < P(F_{ir'}=1) \mbox{ and } P(D_{ir}=1) < P(D_{ir'}=1),
$$
when $r'$ has a larger bundle size or menu size than $r$.
\end{hypothesis}

To refine the picture, we also report results related to our pre-registered secondary hypotheses.\footnote{Appendix Table~\ref{tab:hypothesis_map} lists all pre-registered primary and secondary hypotheses and links them to their empirical test.} We expect dominance violations and score-fit deviations to be distinct behavioral objects with different sources, and dominance violations---being model-free and salient---to be more readily corrected at reconsideration. Specifically, $P(F_{ir}=1\mid D_{ir}=1)$ should exceed $P(F_{ir}=1\mid D_{ir}=0)$, and this comparison should be even more pronounced than the difference between $P(F_{ir}=1\mid M_{ir}=1)$ and  $P(F_{ir}=1\mid M_{ir}=0)$. Finally, we expect calibrated confidence to predict fewer reversals so that it should be higher under \BSD.

\section{Results}
\label{sec:results}

Unless otherwise stated, choice-level regressions use mixed logit specifications with random intercepts for participant and session arm, the set of participants in a session assigned to the same mechanism, whose menus derive from the same live market. Primary tests use the pre-specified Holm correction within families, secondary tests the Benjamini--Hochberg correction.\footnote{Appendix Table~\ref{tab:secondary} collects the secondary hypothesis tests.} We report uncorrected estimates when useful for interpretation, but conclusions follow the corrected tests.

\subsection{Choice inconsistency}
\label{subsec:results-inconsistency}

As already discussed, serial dictatorships provide the most favorable strategic environment for accurate reporting. Hypothesis~\ref{hyp:inconsistency} asks whether this benchmark holds. It does not. 

Across treatments, 17.7 percent of choices do not survive the reconsideration stage, and 18.8 percent of initial choices are objectively dominated (Table~\ref{tab:descriptives-overview}).
In \SSD, participants reverse their initial choice in 17.0 percent of reconsideration opportunities (95\% CI [15.3, 18.7]). In \BSD, the corresponding rate is 18.4 percent ([16.5, 20.3]). Dominance violations are similarly frequent: 18.9 percent of initial choices in \SSD\;([16.7, 21.1]) and 18.8 percent in \BSD \;([16.5, 21.1]) violate objective dominance.

\begin{table}[htbp]
\centering
\captionsetup{width=\linewidth, font=footnotesize}
\caption{Descriptive statistics by mechanism}
\label{tab:descriptives-overview}
\footnotesize
\begin{tabular}{lcccc}
\toprule
 & Overall & Menu SD & Binary SD & $p$ (MWU) \\
\cmidrule(lr){2-2} \cmidrule(lr){3-4} \cmidrule(lr){5-5}
\midrule
Flip rate & \num{0.177} & \num{0.170} & \num{0.184} & \num{0.589} \\
 & (\num{0.147}) & (\num{0.137}) & (\num{0.156}) & \\
Dom.\ violations & \num{0.188} & \num{0.189} & \num{0.188} & \num{0.555} \\
 & (\num{0.180}) & (\num{0.174}) & (\num{0.187}) & \\
Score deviation & \num{0.410} & \num{0.417} & \num{0.403} & \num{0.440} \\
 & (\num{0.228}) & (\num{0.231}) & (\num{0.226}) & \\
Expected profit (EUR) & \num{9.085} & \num{8.743} & \num{9.412} & \num{0.043}\sym{**} \\
 & (\num{3.486}) & (\num{3.580}) & (\num{3.369}) & \\
Realized payoff (EUR) & \num{11.134} & \num{11.369} & \num{10.909} & \num{0.607} \\
 & (\num{9.946}) & (\num{9.926}) & (\num{9.978}) & \\
Confidence & \num{4.719} & \num{4.552} & \num{4.877} & \num{0.005}\sym{***} \\
 & (\num{1.322}) & (\num{1.369}) & (\num{1.259}) & \\
\midrule
$N$ & \num{494} & \num{241} & \num{253} & \\
\bottomrule
\end{tabular}
\captionsetup{width=\linewidth, font=scriptsize}
\caption*{\textit{Notes:} Participant-level means, standard deviations in parentheses. The $p$-value column reports cell-by-cell Mann--Whitney $U$ tests comparing Menu SD vs.\ Binary SD on each outcome; these are treatment-comparison tests, not tests of whether the outcome differs from zero. Asterisks refer to these unadjusted $p$-values. \textit{Dom.\ violations} and \textit{Score deviation} overlap: both include the 7.1 percent of initial choices that are objectively dominated and also a deviation from the elicited score. \textit{Confidence} is the participant's answer to ``How confident are you in the 15 choices you just made?'', given once after the fifteen choices and before the reconsideration stage, on a scale from 1 (not at all) to 7 (very confident). $^{*}$ $p<0.10$, $^{**}$ $p<0.05$, $^{***}$ $p<0.01$.}
\end{table}

\begin{result} \label{result1}
Choices in serial dictatorship are substantially inconsistent. Approximately one in six incentivized choices is reversed upon reconsideration (17.7\%), and almost one in five initial choices violates objective dominance (18.8\%).
\end{result}

These rates are not driven by a few inattentive participants. On the contrary, the vast majority of participants choose inconsistently: 74.3 percent of participants commit at least one dominance violation over the 15 rounds, and 83.8 percent reverse at least one initial choice on reconsideration. At the same time, participants are clearly not choosing at random: observed dominance-violation rates lie roughly 15 percentage points {below} chance in every (mechanism $\times$ complexity) cell,\footnote{See Appendix Table~\ref{tab:ndi}.} so dominated choices cannot be explained by random choice.

Two further observations complete the picture. First, the inconsistency is not an artifact of participants who did not understand the mechanism. After the mechanism instructions, participants answered a three-question comprehension check with two attempts. 382 of the 494 participants passed at the first attempt, 71 passed at the second, and 41 failed both (nobody was excluded). Participants who failed do make more mistakes--- they violate dominance in roughly 27 to 31 percent of rounds, compared with 15 to 17 percent among first-time passers--- but they are too few to account for the phenomenon. First-time passers commit 71 percent of all dominance violations in \SSD\ and 61 percent in \BSD, and between 67 and 77 percent of all reversals, and only 30 percent of them complete the 15 rounds without a single dominance violation. Had the 41 double-failers erred at the passers' rate, the overall dominance-violation rate would be 17.8 rather than 18.8 percent. The mechanism comparison is likewise unchanged within the first-time passers alone.\footnote{See Appendix Table~\ref{tab:comp_mistakes}.}

Second, reversals are not explained by fast clicking. Rounds that are later reversed were not decided faster than rounds that are not: initial choices take 29.6 seconds on reversed rounds and 28.6 seconds on the others, and the within-participant difference is positive ($+3$ percent, $p=0.018$). On the revisit page itself, reversed rounds take about 6 percent longer than non-reversed ones ($p<0.001$), and this holds under both mechanisms. At the participant level, frequent flippers are indistinguishable from infrequent ones: mean initial-choice time is flat across quartiles of the individual reversal rate (29.3, 28.2, 29.0 and 28.7 seconds; Kruskal--Wallis $p=0.83$), and so is revisit-page time.\footnote{See Appendix Tables~\ref{tab:flip_times} and~\ref{tab:nf_times}.} Reversals therefore look like reconsideration rather than inattention. 

\begin{table}[htbp]
\centering
\captionsetup{width=\linewidth, font=footnotesize}
\caption{Inconsistencies in initial choices and reconsideration flip rates}
\label{tab:errortypes-shares}
\footnotesize
\begin{tabular}{lcc}
\toprule
 & Share of initial choices & Flip rate \\
\midrule
No measured inconsistency & \num{0.472} & \num{0.089} \\
Dominance violation & \num{0.188} & \num{0.364} \\
Score inconsistency & \num{0.410} & \num{0.208} \\
\bottomrule
\end{tabular}
\captionsetup{width=\linewidth, font=scriptsize}
\caption*{\textit{Notes:} All 7,410 initial choices, pooled over both mechanisms. A choice is a dominance violation if it is objectively dominated by another bundle in the menu, and a score deviation if it is not a top-scoring bundle under the participant's elicited score. The two error rows overlap: choices that are both (a share of \num{0.071}) are counted in each, so the shares do not sum to one. \textit{Flip rate} is the share of the initial choices in the row that were reversed at reconsideration.}
\end{table}

A flip is not necessarily a choice mistake as it could reflect indifference, stochastic tastes, or incomplete preferences \citep{Gerasimou2018, AgranovOrtoleva2017}. 
However, our data in Table~\ref{tab:errortypes-shares} do not suggest that our participants are largely indifferent over bundles. 47.2 percent of choices at the mechanism stage are neither dominated nor in conflict with the score ranking, and only 8.9 percent of such choices are reversed. 
Opposed to that, we find that 20.8 percent of score-inconsistent choices are reversed. Moreover, a mixed-logit specification shows a significantly positive score gain through flipping,\footnote{See Appendix Table~\ref{tab:gain}.} which again suggests that a flip is a correction rather than a randomization: participants use the reconsideration stage to reduce score inconsistency.
Even stronger is the evidence regarding dominated choices, which are flipped far more often than non-dominated ones (36.4\% vs.\ 13.4\%; $p<0.001$), in line with our interpretation that these choices are indeed considered mistakes.
When the alternative offered at revisit itself dominates the initial choice, the flip rate rises to 51.7\%, from 24.5\% when it does not.\footnote{See Appendix Table~\ref{tab:flip_2x2}.}

Our preregistered hypothesis on internal consistency states that participants' actual choices are strongly correlated with their elicited score ranking. At the participant level, however, there is no association between a participant's flip rate and their score-model deviation rate (Spearman $\rho=-0.01$, $p=0.78$; Table~\ref{tab:primary}). One reason is that the elicited score ignores difficulty: bundles that differ only in difficulty have the same score, and choosing the harder twin counts as consistent with the model. Breaking such ties in favor of the easier bundle, which amounts to counting every dominated choice as a deviation, reclassifies 12 percent of choices and turns the correlation positive ($\rho=0.19$, $p<0.001$).\footnote{See Appendix Table~\ref{tab:errortypes-tiebreak}.}

\subsection{Binary decomposition: does it help, and for whom? }
\label{subsec:results-mechanism-heterogeneity}

 Hypothesis~\ref{hyp:mechanism} asks whether decomposing the menu into binary comparisons improves choice quality on average. 
Motivated by \citet{CaspariKhanna2024}, our idea was simplification: since a binary comparison is simpler than studying a larger menu, a sequence of binary comparisons should lead to better choice quality. However, one can also argue for the opposite, fragmentation: if every decision involves some noise, splitting one decision into several smaller decisions by construction amplifies the noise so that the resulting decision is even more distorted.

On average, the evidence rejects both predictions.
Flip rates are slightly higher in \BSD\ than in \SSD, 18.4 percent versus 17.0 percent, but the difference is small and statistically insignificant. Dominance violations are essentially identical across mechanisms, 18.8 percent in \BSD\ and 18.9 percent in \SSD\ (Table~\ref{tab:descriptives-overview}). Realized payoffs are also statistically indistinguishable across mechanisms (Table~\ref{tab:descriptives-overview}), and so is the score fit of the final bundle, the elicited score of the bundle a participant ends up with (12.8 vs.\ 13.7 points; MWU $p=0.16$; Table~\ref{tab:primary}). Raw expected profit is higher under \BSD, $+0.67$~EUR per participant, and this is the one choice-quality or payoff outcome for which the unconditional difference between mechanisms reaches conventional significance (MWU $p=0.043$; Table~\ref{tab:descriptives-overview}). Expected profit is almost entirely determined by learning-stage ability, and the \BSD{} arm is the higher-ability arm; conditioning on ability, the \BSD{} coefficient is negative ($-0.57$~EUR at mean ability, $p<0.001$; Table~\ref{tab:ame}, Panel~B). The raw difference therefore does not show that binary decomposition raises expected profit.

The effect sizes are correspondingly small. Cliff's delta is $-0.028$ for flip rates and $0.030$ for dominance violations, both classified as negligible.\footnote{Table~\ref{tab:effects}. Cliff's delta is the probability that a randomly drawn Menu SD participant has a higher value of the outcome than a randomly drawn Binary SD participant, minus the probability of the reverse. It ranges from $-1$ to $1$, equals zero when the two distributions overlap completely, and a negative value means the outcome tends to be higher under \BSD. Following \citet{romano2006}, absolute values below $0.147$ are classified as negligible, below $0.33$ as small, and below $0.474$ as medium.} Mixed-logit specifications with random intercepts for participant and session arm deliver a similar conclusion: evaluated at mean ability, the \BSD\ coefficient on dominance violations is small and insignificant, and the coefficient on flips is positive but significant only at the ten percent level (Table~\ref{tab:ame}, Panel~A, ``\BSD'' row). Neither estimate shows the improvement that simplification predicts; both are slightly positive, which the analysis by ability below explains.

\paragraph{Ability.}
As discussed below Hypothesis \ref{hyp:mechanism}, we expected only a minor effect on high-ability participants, but believed that low-ability participants gain from simplification. While \BSD{} indeed has only an insignificant positive impact on the choice quality of high-ability subjects, the effect on low-ability subjects is significant, but in the opposite direction: fragmentation dominates, and decomposing the menu into binary comparisons reduces the choice quality of low-ability participants.

The continuous-ability specification estimates a mixed-logit model for reconsideration flips with an interaction between \BSD\ and the learning-stage ability index (Table~\ref{tab:ame}, Panel~A, column~1). The interaction is negative and statistically significant:
\[
\hat{\beta}_{\BSD \times \text{Ability}}=-0.744,
\qquad
z=-4.57,
\qquad
p<0.001,
\]
with a Benjamini--Hochberg corrected value of \(q<0.001\). Thus, the relative effect of \BSD\ becomes more favorable as measured ability increases. The same pattern holds for objective mistakes: the \BSD~$\times$~Ability interaction on dominance violations is of almost identical size and also significant ($\hat{\beta}=-0.744$, $p<0.01$; Table~\ref{tab:ame}, Panel~A, column~2), so ability matters beyond reconsiderations. For score deviations, the interaction is also negative but indistinguishable from zero. No mixed logit is estimated for that outcome, so the comparison must be drawn in the participant-level linear specification, in which the corresponding flip and dominance interactions are $-0.108$ and $-0.083$ and the score-deviation interaction is $-0.026$ ($p=0.461$).\footnote{See Appendix Table~\ref{tab:e8-continuous-ability}.} Panel~B reveals a welfare-side counterpart: the \BSD\ effect on expected profit grows more positive with ability ($\hat{\beta}_{\BSD \times \text{Ability}}=3.110$, $p<0.001$), consistent with the choice-quality pattern, while the interaction with realized payoff is indistinguishable from zero.

\begin{table}[htbp]
\centering
\captionsetup{width=\linewidth, font=footnotesize}
\caption{Mechanism effects on choice-quality and welfare outcomes (Binary SD $\times$ Ability interaction)}
\label{tab:ame}
\footnotesize
\begin{tabular}{lcc}
\toprule
\multicolumn{3}{c}{Dependent variable: choice-quality and welfare outcomes} \\
\midrule
\multicolumn{3}{l}{\textit{Panel A: Choice-level mixed logit (random intercepts: participant, session arm)}} \\
\addlinespace[2pt]
 & Flip & Dom.\ viol. \\
 & (1) & (2) \\
\midrule
Binary SD & \num{0.182}\sym{*} & \num{0.099} \\
 & (\num{0.103}) & (\num{0.128}) \\
\addlinespace
Binary SD $\times$ Ability & \num{-0.744}\sym{***} & \num{-0.744}\sym{***} \\
 & (\num{0.163}) & (\num{0.214}) \\
\addlinespace
Ability & \num{-0.105} & \num{-0.356}\sym{***} \\
 & (\num{0.077}) & (\num{0.093}) \\
\midrule
$N$ choices    & 7,410 & 7,410 \\
$N$ participants & 494 & 494 \\
\addlinespace
\multicolumn{3}{l}{\textit{Panel B: Participant-level OLS, HC1 robust SE}} \\
\addlinespace[2pt]
 & E[profit] (EUR) & Realized (EUR) \\
 & (3) & (4) \\
\midrule
Binary SD & \num{-0.566}\sym{***} & \num{-1.428} \\
 & (\num{0.157}) & (\num{0.888}) \\
\addlinespace
Binary SD $\times$ Ability & \num{3.110}\sym{***} & \num{0.453} \\
 & (\num{0.246}) & (\num{1.470}) \\
\addlinespace
Ability & \num{3.418}\sym{***} & \num{3.648}\sym{***} \\
 & (\num{0.141}) & (\num{0.543}) \\
\midrule
$N$ participants & 494 & 494 \\
$R^2$            & \num{0.765} & \num{0.073} \\
\bottomrule
\end{tabular}
\captionsetup{width=\linewidth, font=scriptsize}
\caption*{\textit{Notes:} Panel A: choice-level mixed logit with random intercepts for participant and session arm, fit on the binary outcome indicated by the column header. Panel B: participant-level OLS with HC1 heteroskedasticity-robust standard errors. Both panels include the Binary SD $\times$ Ability interaction. Standard errors in parentheses. Ability is centered at its sample mean (\num{-0.003}), so the Binary SD row reads as the Binary SD effect at mean ability; the Binary SD $\times$ Ability row gives the change in that effect per unit of the ability index. Columns flagged with `singular' in the rendered log fell back to a participant-only random-intercept specification. The Panel A flip-column interaction is the preregistered S5 test. $^{*}$ $p<0.10$, $^{**}$ $p<0.05$, $^{***}$ $p<0.01$.}
\end{table}

For interpretation, Table~\ref{tab:ability} reports the corresponding median-split estimates.\footnote{Appendix Table~\ref{tab:ability-combined} extends the split to all five preregistered outcomes.} Among low-ability participants, \BSD \; raises the flip rate from 17.2 percent to 23.5 percent ($p=0.006$). This is a large deterioration in a setting where the mechanism has already removed strategic uncertainty. Among high-ability participants, flip rates are lower under \BSD, 14.1 percent compared with 16.8 percent under \SSD, though this difference is not statistically significant ($p=0.167$).

\begin{table}[htbp]
\centering
\captionsetup{width=\linewidth, font=footnotesize}
\caption{Flip rate by ability group and mechanism}
\label{tab:ability}
\footnotesize
\begin{tabular}{lccccc}
\toprule
& Menu SD & Binary SD & MW $p$ & $N$ (Menu SD) & $N$ (Binary SD) \\
\midrule
Low ability & \num{0.172} & \num{0.235}\sym{***} & \num{0.006} & \num{131} & \num{116} \\
& (\num{0.129}) & (\num{0.171}) & & & \\
High ability & \num{0.168} & \num{0.141} & \num{0.167} & \num{110} & \num{137} \\
& (\num{0.147}) & (\num{0.128}) & & & \\
\bottomrule
\end{tabular}
\captionsetup{width=\linewidth, font=scriptsize}
\caption*{\textit{Notes:} Standard deviations in parentheses. Median split on overall ability. Asterisks on the Binary SD column indicate the preregistered within-stratum Mann--Whitney test of equal flip-rate distribution under Menu SD vs.\ Binary SD; the corresponding $p$-value is reported in the MW $p$ column. The heterogeneity claim that the Binary SD effect \emph{differs} between low- and high-ability participants is the Binary SD $\times$ Ability interaction reported in Table~\ref{tab:ame} and in the text. $^{*}$ $p<0.10$, $^{**}$ $p<0.05$, $^{***}$ $p<0.01$.}
\end{table}

\begin{result} 
Binary decomposition significantly hurts low-ability participants, while it directionally benefits high-ability ones.
\end{result}

The ability imbalance reported in Section~\ref{sec:design} cannot generate this result,
because it would make \BSD{}  look better. Ability is measured in the learning stage, before market
assignment, so it is a pre-treatment covariate: the \BSD{} arm is the higher-ability arm,  and higher ability predicts fewer reversals. Any bias from the imbalance therefore makes
\BSD{} look better than it is, and the imbalance reproduces \emph{within} the
low-ability stratum, where \BSD{} participants are the stronger group. Consistent with
this, controlling for the continuous ability index inside the low-ability stratum
\emph{increases} the estimated harm from $+0.063$ to $+0.071$ ($p<0.001$), and splitting
ability within rather than across arms, which removes the compositional difference by
construction, leaves it at $+0.056$ ($p=0.004$). The direction and significance of the ability result are unchanged under every treatment of the imbalance.\footnote{See Appendix~\ref{app:robustness}.}

\begin{figure}[h!]
\centering
\includegraphics[width=0.9\linewidth]{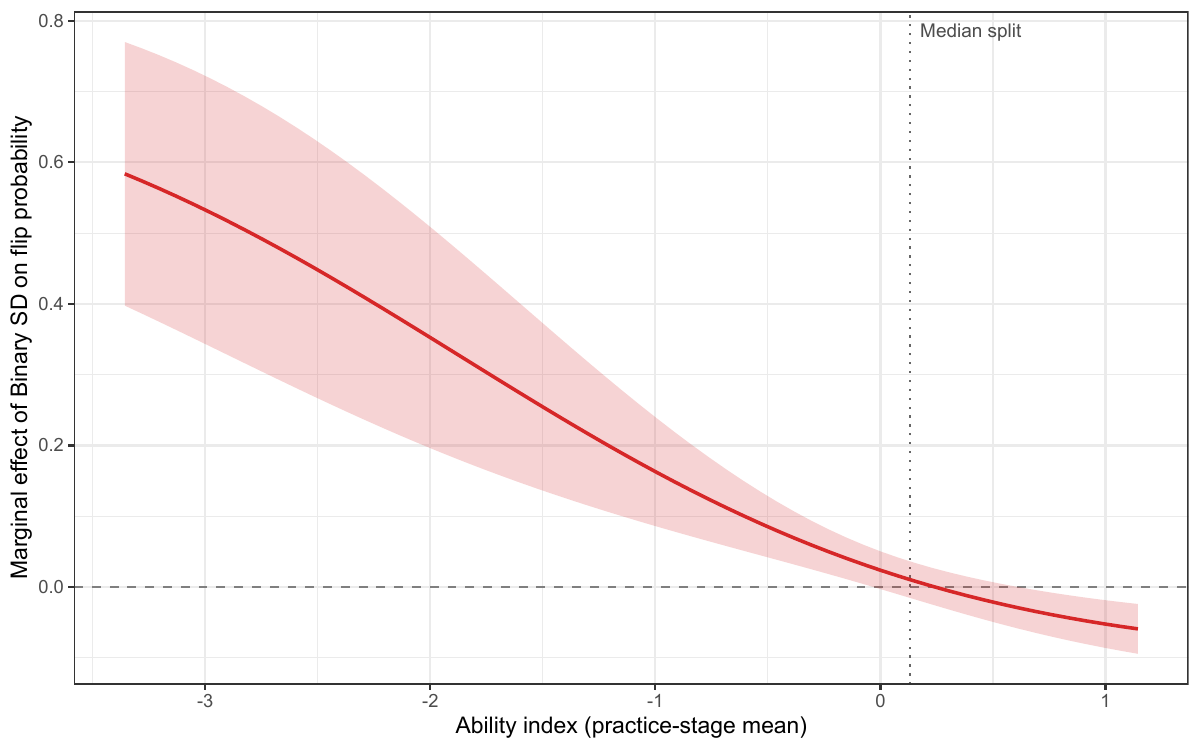}
\caption{The effect of binary decomposition by participant ability}
\label{fig:ability_margins}
\captionsetup{width=\linewidth, font=scriptsize}
\caption*{\textit{Notes:} Marginal effect of \BSD\ on the choice-level flip probability across the practice-stage ability index, computed from the mixed-logit specification $\text{flip}_{ij} \sim \text{mechanism} \times \text{ability}_i + (1\,|\,\text{participant}_i) + (1\,|\,\text{session arm}_{s(i)})$. The shaded band is the delta-method 95\% confidence interval based on the fixed-effect covariance matrix. The dotted vertical line marks the sample median of the ability index used for the median split in Table~\ref{tab:ability}. The marginal effect is positive (worse choice consistency under \BSD) for low-ability participants, crosses zero near the median, and turns slightly negative for high-ability participants.}
\end{figure}

Table~\ref{tab:e4b2-cells} decomposes the treatment effect by ability group and bundle complexity. A likelihood-ratio test does not reject the null of no three-way interaction ($\chi^2(2)$ = 2.82, p = 0.244), so the cell-level contrasts below are descriptive. Within that limit, two cells survive multiple-testing correction across the six comparisons in the table. High-ability participants benefit from \BSD\ in the low-complexity (one-task) rounds, where flip rates fall from 24.4 percent to 15.5 percent (Holm $p=0.012$). In medium and high-complexity rounds, their treatment differences are small and insignificant. Low-ability participants have higher flip rates under \BSD\ at every complexity level, significantly so in medium rounds, where the flip rate rises from 13.9 to 22.6 percent (Holm $p=0.026$).\footnote{The complexity blocks are played in the fixed Low--Medium--High order, so the strong high-ability Low-block effect coincides with the first complexity block. We cannot rule out that part of the high-ability gain from \BSD\ in the Low block reflects an early-experiment effect rather than a complexity-specific mechanism. The asymmetry between low- and high-ability participants within the same block, however, is not subject to this concern.}

\begin{table}[htbp]
\centering
\captionsetup{width=\linewidth, font=footnotesize}
\caption{Flip rate by ability $\times$ mechanism $\times$ complexity}
\label{tab:e4b2-cells}
\footnotesize
\begin{tabular}{lccccc}
\toprule
Complexity & Menu SD & Binary SD & $p$ (MWU) & $p$ (Holm) & MDE \\
\midrule
\multicolumn{6}{l}{\textbf{Low ability}}\\
Low & \num{0.218} (\num{0.016}) & \num{0.267} (\num{0.018}) & \num{0.085} & \num{0.340} & \num{0.082} \\
Medium & \num{0.139} (\num{0.014}) & \num{0.226} (\num{0.017}) & \num{0.005} & \num{0.026}\sym{**} & \num{0.076} \\
High & \num{0.159} (\num{0.014}) & \num{0.212} (\num{0.017}) & \num{0.123} & \num{0.368} & \num{0.074} \\
\multicolumn{6}{l}{\textbf{High ability}}\\
Low & \num{0.244} (\num{0.018}) & \num{0.155} (\num{0.014}) & \num{0.002} & \num{0.012}\sym{**} & \num{0.073} \\
Medium & \num{0.129} (\num{0.014}) & \num{0.139} (\num{0.013}) & \num{0.492} & \num{0.985} & \num{0.065} \\
High & \num{0.133} (\num{0.014}) & \num{0.128} (\num{0.013}) & \num{0.867} & \num{0.985} & \num{0.064} \\
\bottomrule
\end{tabular}
\captionsetup{width=\linewidth, font=scriptsize}
\caption*{\textit{Notes:} Cell means with standard errors in parentheses. Within-cell Mann--Whitney $U$ test of Menu SD vs.\ Binary SD on participant-level mean flip rates inside each (ability, complexity) cell. $p$ (Holm) applies a Holm correction to the six within-cell tests; asterisks reflect the Holm-corrected $p$-value. The omnibus three-way interaction test is reported in the text. MDE is the minimum detectable difference in participant-level mean flip rates at 80 percent power and a two-sided 5 percent level, $(z_{0.975}+z_{0.80})\,s_p\sqrt{1/n_S+1/n_B}$, with $s_p$ the pooled standard deviation of participant-level flip rates in the cell and $n_S$, $n_B$ the participants per arm (low ability: 131 Menu SD, 116 Binary SD; high ability: 110 and 137). $^{*}$ $p<0.10$, $^{**}$ $p<0.05$, $^{***}$ $p<0.01$.}
\end{table}

This pattern suggests that binary decomposition changes how participants process the choice problem. For high-ability participants, the binary format can discipline comparisons when the underlying alternatives are easy to evaluate. For low-ability participants, however, the same format fragments the decision problem. The participant may make each local comparison with apparent ease, but the sequence does not aggregate into a stable global choice.
In contrast to what we expected, the negligible effect of binary decomposition among high-ability subjects is not driven by them already having a high choice quality in \SSD. For each complexity level, flip rates are roughly similar across ability levels in \SSD, but binary decomposition increases flip rates only for low-ability participants.

\subsection{Which complexity margin matters?}
\label{subsec:results-complexity}

Hypothesis~\ref{hyp:complexity} points t two sources of cognitive difficulty that our design varies independently: bundle dimensionality (one, two, or three tasks per bundle) and menu size (which shrinks from seven to three within each block). One-task rounds, the simplest bundles in the design, exhibit the highest flip rates of any complexity block, so if any of the two margins matters, it is not bundle dimensionality. The evidence points to menu size instead, although not %in the form of a gradient.
monotonically.

The predicted monotonic relationship between bundle dimensionality and inconsistency does not hold. The pattern runs the other way: in the round-level regressions, both the Medium and High complexity blocks have significantly lower flip rates than the one-task block (Table~\ref{tab:main-results}).\footnote{The Medium${-}$Low and High${-}$Low coefficients in the flip column are both $-0.442$ with the same standard error. 
The estimates are not in fact equal, they differ in the eighth significant figure and the data underneath them are not symmetric: flips are distributed differently across menu sizes in the two blocks, and 274 of the 494 participants flip a different number of times in Medium than in High.} Flip rates do not increase from one-task to three-task bundles: in both mechanisms the one-task ``Low'' block exhibits the highest flip rate (Table~\ref{tab:complexity}), and dominance violations exhibit the same lack of monotonicity.\footnote{Raw dominance-violation rates pool a behavioral component with a mechanical one: the share of bundles in each menu that are dominated by some other menu bundle varies by complexity block as a combinatoric property of the bundle space. Normalizing the observed dominance-violation rate against this menu-specific baseline (Appendix Table~\ref{tab:ndi}) yields excess incidence of roughly $-15$ percentage points in every (mechanism $\times$ complexity) cell. Participants pick dominated bundles much less often than chance, but to the same degree everywhere. The mechanism difference in normalized incidence is null ($p=0.455$) and the complexity gradient is small and in the direction of more, not less, avoidance with complexity ($-0.016$/step, $p=0.047$).} The data do not support the view that inconsistencies are primarily driven by the number of attributes inside each bundle.

By contrast, the menu-size coefficient in the round-level flip regression is positive and highly significant (Table~\ref{tab:main-results}).\footnote{Neither margin was order-randomized: complexity blocks are played in the fixed Low--Medium--High sequence, and within each block menu size declines from seven (rank 1) to three (rank 5) as ranks are realized. Bundle complexity is thus confounded with block order, and menu size with within-block position. The asymmetry between the two margins is harder to attribute to order effects than each margin is on its own: if order effects drove the menu-size association, an analogous trend should appear across complexity blocks, which it does not.} However, menu size does not have a monotonic effect and only matters through a step between three and four bundles. The descriptive rates show what this coefficient summarizes. In both mechanisms, flips are least frequent in the three-bundle menu (13.4 percent under \SSD\ and 13.3 percent under \BSD) and higher in every larger menu (between 17.0 and 19.4 percent under \SSD, and between 17.7 and 21.7 percent under \BSD), with no systematic further increase from four to seven bundles (Table~\ref{tab:mechxmenu}). The linear coefficient should therefore be read as a summary of the step between three and four bundles, not as a constant marginal effect of each additional bundle. The within-participant evidence is in line with this: the mean within-participant Kendall's $\tau$ between menu size and flip rate is $+0.060$ (SD $0.350$), and is strictly positive for 52.4 percent of the participants for whom it is defined (217 of 414; one-sample $t$-test against zero, one-sided $p<0.001$).\footnote{The statistic is undefined for the 80 participants who never reversed a choice, because flipping is constant for them, so the denominator is the 414 participants with at least one reversal rather than the full sample; 217 is 43.9 percent of all 494 participants. Assigning $\tau=0$ to those 80 leaves the conclusion intact (mean $+0.050$, $p<0.001$). Note also that $\tau$ is computed within each complexity block across the five menu sizes and then averaged over the blocks in which it is defined, which is all three blocks for 109 participants, two for 151, and one for 154.} Because $\tau$ is a rank correlation over all five menu sizes, a single step at the smallest menu is enough to make it positive, so it does not establish a gradient either. Reversals are least frequent when participants choose from three bundles. Beyond that, larger feasible sets are not associated with more frequent reversals.

Decision times show a different pattern. Within participant, the time spent on the initial choice rises steadily with menu size in both mechanisms, from about 22 seconds at three bundles to 31 seconds at six under \SSD\ and to 40 seconds at seven under \BSD, and, unlike reversal rates, it keeps rising beyond four bundles (Wald test of equal effects for sizes four to seven, $p<0.001$).\footnote{See Appendix Table~\ref{tab:size_time}. Three qualifications apply. Menu size equals eight minus the priority rank, so it is collinear with within-block position; under \BSD\ a menu of $k$ bundles requires $k-1$ clicks, so part of its gradient is mechanical; and the seven-bundle page is the only choice page in a block not preceded by a group wait page, which lowers its duration under \SSD.} Bundle dimensionality, by contrast, does not raise decision time: controlling for menu size, two- and three-task rounds take less time than one-task rounds under \SSD, and under \BSD\ only the three-task block takes longer. Larger menus therefore cost time as well as consistency, whereas more complex bundles cost little of either.

\begin{table}[htbp]
\centering
\captionsetup{width=\linewidth, font=footnotesize}
\caption{Main treatment effects: random-intercepts logits (round-level)}
\label{tab:main-results}
\footnotesize

\begin{tabular}{lcc}
\toprule
& Flip rate & Dom.\ violations \\
\midrule
Menu SD (marg.\ prob.) & 0.144 & 0.139 \\
Binary SD (marg.\ prob.) & 0.156 & 0.131 \\
\addlinespace
Binary SD & +0.092 & $-$0.068 \\
\quad (SE) & (0.104) & (0.142) \\
Menu size & +0.085*** & +0.537*** \\
\quad (SE) & (0.022) & (0.026) \\
Medium (vs Low) & $-$0.442*** & $-$0.062 \\
\quad (SE) & (0.076) & (0.082) \\
High (vs Low) & $-$0.442*** & $-$0.122 \\
\quad (SE) & (0.076) & (0.082) \\
\addlinespace
Num.\ obs.\ & \num{7410} & \num{7410} \\
\bottomrule
\end{tabular}

\captionsetup{width=\linewidth, font=scriptsize}
\caption*{\textit{Notes:} Random-intercepts logistic regressions (random intercepts for participant and session arm) of binary round-level outcomes on a Binary SD dummy, menu size ($\in \{3,\dots,7\}$, numeric), and complexity as an unordered factor with Low as the reference category (so the reported coefficients are Medium${-}$Low and High${-}$Low log-odds contrasts). Coefficients are on the log-odds scale. The first two rows report population-average marginal probabilities, evaluated by averaging predicted probabilities across the estimation sample with the Binary SD dummy fixed to 0 and 1 respectively (random effects set to zero). $^{*}$ $p<0.10$, $^{**}$ $p<0.05$, $^{***}$ $p<0.01$.}
\end{table}

\begin{result}
The relevant complexity margin is the number of alternatives in the menu, not the number of task components in the bundle. Reversals are least frequent when participants choose from the smallest, three-bundle menu and do not rise further from four to seven bundles, whereas integrating several task components within a bundle does not increase choice inconsistency.
\end{result}

\subsection{Confidence}
\label{subsec:results-confidence}

If binary decomposition makes each local comparison feel easier, participants may report higher confidence under \BSD\ even when their choices are no more reliable.
Table~\ref{tab:e5a-conf-ability} reports confidence and happiness by treatment and ability group, and it confirms this idea. Overall, participants in \BSD\ report higher confidence than participants in \SSD. The Mann--Whitney test gives \(p=0.005\), with a Benjamini--Hochberg corrected value of \(q=0.015\). The within-stratum pattern is consistent with the increase being concentrated among low-ability participants: their mean confidence rises from 4.59 under \SSD\ to 5.06 under \BSD\ ($p=0.009$), while for high-ability participants the difference is smaller (4.51 vs.\ 4.72) and only marginally significant ($p=0.095$). A formal participant-level interaction of mechanism with the \emph{continuous} ability index confirms this pattern: the \BSD~$\times$~Ability coefficient on confidence is negative and significant (\(\hat{\beta}=-0.640\), \(p=0.002\)), meaning that the confidence boost from \BSD\ shrinks as ability rises, i.e., it is concentrated among low-ability participants.\footnote{See Appendix Table~\ref{tab:e8-continuous-ability}. The corresponding median-split specification ($\text{mechanism} \times \mathbf{1}\{\text{ability}=\text{Low}\}$) gives an interaction of $+0.259$ ($p=0.274$), so the concentration among low-ability participants is established by the continuous specification rather than by the coarser split. The descriptive within-stratum differences are visible in Table~\ref{tab:e5a-conf-ability}.} Happiness does not differ significantly across mechanisms.

\begin{table}[htbp]
\centering
\captionsetup{width=\linewidth, font=footnotesize}
\caption{Confidence and happiness by ability group and mechanism}
\label{tab:e5a-conf-ability}
\footnotesize
\begin{tabular}{lcccc}
\toprule
& Menu SD & Binary SD & $p$ (MWU) & Cliff's $\delta$ \\
\midrule
\multicolumn{5}{l}{\textbf{Low ability}}\\
Confidence & \num{4.588} & \num{5.060} & \num{0.009}\sym{***} & \num{-0.188} \\
& (\num{1.451}) & (\num{1.404}) & & \\
Happiness & \num{4.672} & \num{4.819} & \num{0.668} & \num{-0.031} \\
& (\num{1.526}) & (\num{1.289}) & & \\
$N$ & \num{131} & \num{116} & & \\
\multicolumn{5}{l}{\textbf{High ability}}\\
Confidence & \num{4.509} & \num{4.723} & \num{0.095}\sym{*} & \num{-0.118} \\
& (\num{1.269}) & (\num{1.103}) & & \\
Happiness & \num{4.555} & \num{4.599} & \num{0.669} & \num{-0.031} \\
& (\num{1.268}) & (\num{1.263}) & & \\
$N$ & \num{110} & \num{137} & & \\
\bottomrule
\end{tabular}
\captionsetup{width=\linewidth, font=scriptsize}
\caption*{\textit{Notes:} Standard deviations in parentheses below the corresponding means. $^{*}$ $p<0.10$, $^{**}$ $p<0.05$, $^{***}$ $p<0.01$.}
\end{table}

\begin{result}
\BSD{} raises reported confidence.
This effect is mostly driven by low-ability participants, the group whose choices become less consistent. 
\end{result}

The confidence pattern therefore does not corroborate the choice-quality results. Among high-ability participants, where choice quality is essentially unchanged across mechanisms, confidence is essentially unchanged, too. Among low-ability participants, choice quality deteriorates under \BSD\ while confidence rises: the procedure feels easier without producing more reliable choices.

The confidence effect is not a by-product of quicker decisions. Participants under \BSD\ spend about five seconds longer per round than under \SSD, in both ability groups,\footnote{See Appendix Table~\ref{tab:time-combined}.} so the binary format makes the decision feel easier while taking more time, and the additional time does not translate into more reliable choices for low-ability participants. Nor is the residual inconsistency under \BSD\ explained by confidence itself: controlling for reported confidence, low-ability participants still flip $+0.075$ more often per round under \BSD\ than under \SSD\ ($p<0.001$, HC1), while high-ability participants show no analogous shift ($-0.025$, $p=0.171$). The binary format changes the perceived difficulty of the decision without changing its underlying reliability.

\begin{figure}[!t]
\centering
\includegraphics[width=\linewidth]{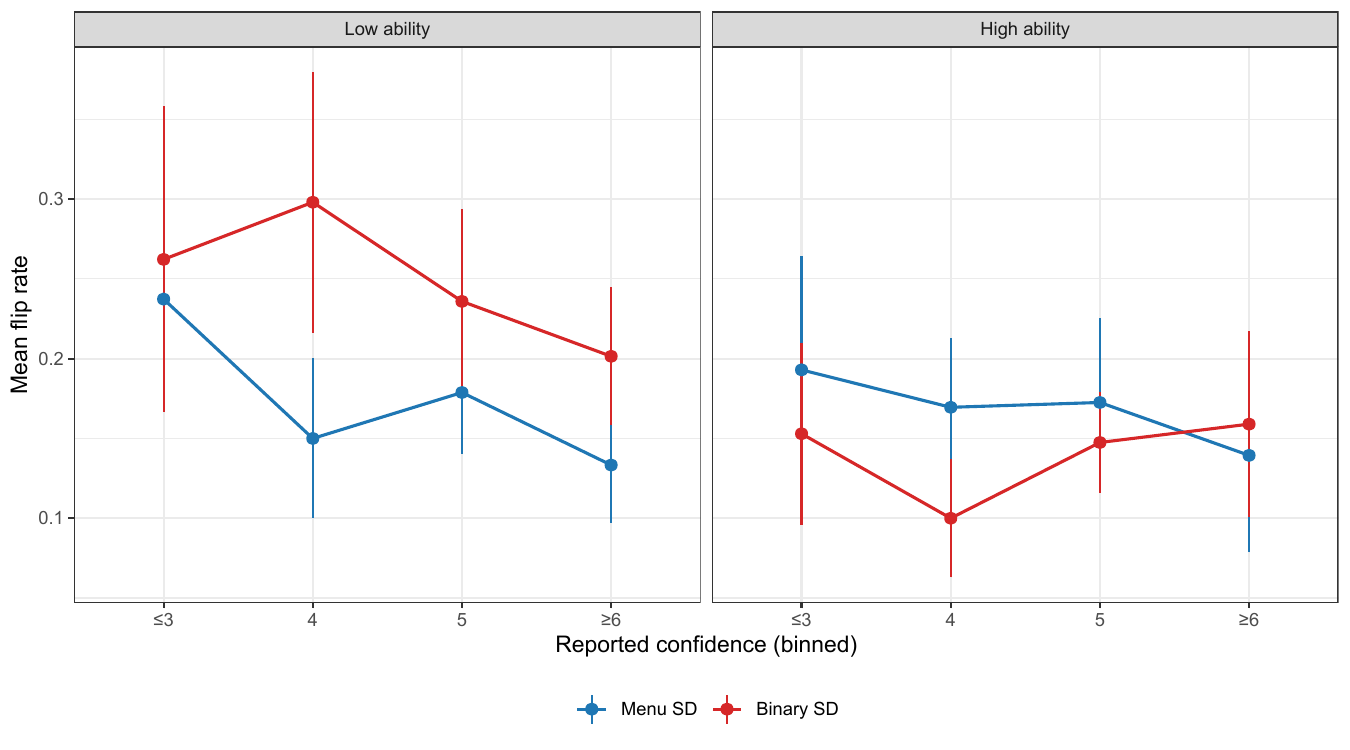}
\caption{Flip rate by reported confidence, ability and mechanism}
\label{fig:calibration}
\captionsetup{width=\linewidth, font=scriptsize}
\caption*{\textit{Notes:} Participant-mean flip rate by reported confidence bin, faceted by mechanism and ability group. Confidence is the post-decision self-report on a 1--7 scale, binned into $\leq 3$, $4$, $5$, $\geq 6$. Markers are means across participants in each bin; vertical bars are normal-approximation 95\% confidence intervals. Bins with fewer than five participants are suppressed. For high-ability participants, reported confidence predicts fewer flips and the two mechanisms overlap. For low-ability participants, the \BSD\ line lies above the \SSD\ line at every confidence level: the same group whose subjective confidence rises under \BSD\ (Table~\ref{tab:e5a-conf-ability}) flips more often regardless of how confident they report being.}
\end{figure}

\section{Discussion}
\label{sec:discussion}

Based on \cite{CaspariKhanna2024}, our central hypothesis was that binary decomposition increases choice quality through simplification of the choice process. Our evidence does not support this hypothesis: binary decomposition does not improve choice quality on average and increases reversals among low-ability participants.
Moreover, augmenting choice complexity through bundling tasks did not increase choice inconsistency. Surprisingly, we observe more flips in low-complexity environments.
In this section, we discuss what may have led to this divergence from our hypotheses.

Adopting an Occam’s razor framework, questioning our initial assumption that more complex choices necessarily produce more choice mistakes might explain both observations. For example, \cite{kahneman2011book} popularized dividing thinking into System 1 (thinking fast) and System 2 (thinking slow), which proposes that people think deeper in complex environments, but only decide intuitively in less complex environments. Under this view, one-task bundles invite intuitive System-1 choices that are more easily reversed, whereas multi-task bundles trigger deliberate System-2 thinking. Similarly, this theory suggests that people think harder when selecting from a menu compared to a seemingly easy binary decision. The response-time data offer little support for this account, however. Participants spend more time per round under \BSD\ than under \SSD\ (Table~\ref{tab:time-combined}), one-task rounds are not decided faster than multi-task rounds under \SSD\ (Table~\ref{tab:size_time}), and choices that are later reversed were not made faster than those that are kept (Table~\ref{tab:flip_times}).

Alternatively, random-choice and random-preference models provide a theoretical foundation of fragmentation. In the former, the decision maker trembles with a small probability at each decision node, and, in the latter, a mean-zero noise term is added to all bundle utilities, which can sometimes overturn preference rankings. In both models, there is more decision noise in \BSD\ by construction because the noise is compounded rather than occurring just once. Note that one noise-driven decision can cascade as a rejected alternative never resurfaces again. However, this theory predicts that flip rates in \BSD\ increase monotonically in the number of binary choices, which is not the case as Table \ref{tab:mechxmenu} documents.

Finally, our treatment conditions juxtapose the strongly obviously strategy-proof \SSD\ with \BSD\ which is only weakly obviously strategy-proof. 
Conditional on reaching a decision node in \SSD, the actions of others are irrelevant so that the own decision pins down the best and worst case outcome. Hence, not selecting a favorite bundle is obviously dominated and this obvious dominance is strong because each player moves exactly once.
\BSD{} is obviously strategy-proof for the same reason, but not strongly so because the fragmentation implies that a subject has to reason about her own future decisions.
In this light, strategic confusion of low-ability participants in \BSD\ also offers a potential explanation for our observations.

\section{Conclusion}
\label{sec:conclusion}

While a large part of the market design literature focuses on how to incentivize the truthful revelation of preferences, our research question comes one step earlier: Can people identify what they prefer, and does the choice format help them?  
In sequential serial dictatorships, participants simply pick their favorite  among the offered options. Even in this strategically trivial environment, we show that a substantial share of choices is reversed on reconsideration or objectively dominated. We therefore attribute choice inconsistencies to the choice process rather than to strategic considerations.

By comparing two sequential serial dictatorships, one in which the participants select task bundles from a full menu and one in which this menu is decomposed into a sequence of pairwise comparisons, we isolate whether making each step locally simpler leads to better final choices.
On average, choice inconsistencies occur with the same frequency in both implementations. However, this observation hides the opposing effects across ability groups. While high-ability participants benefit modestly from binary decomposition, the choice quality of low-ability participants decreases due to fragmentation. Nevertheless, this group reports a higher confidence in their choices. Finally, choice inconsistency rises with the number of bundles in the menu, but not with the number of tasks within a bundle. To the best of our knowledge, we are the first to distinguish these forms of choice complexity in market design.

\bibliographystyle{aer}
\bibliography{references}

@article{chen2016otree,
  title={oTree—An open-source platform for laboratory, online, and field experiments},
  author={Chen, Daniel L and Schonger, Martin and Wickens, Chris},
  journal={Journal of Behavioral and Experimental Finance},
  volume={9},
  pages={88--97},
  year={2016},
  publisher={Elsevier}
}

@article{Li2017,
  author  = {Li, Shengwu},
  title   = {Obviously Strategy-Proof Mechanisms},
  journal = {American Economic Review},
  year    = {2017},
  volume  = {107},
  number  = {11},
  pages   = {3257--3287}
}

@article{PyciaTroyan2023,
  author  = {Pycia, Marek and Troyan, Peter},
  title   = {A Theory of Simplicity in Games and Mechanism Design},
  journal = {Econometrica},
  year    = {2023},
  volume  = {91},
  number  = {4},
  pages   = {1495--1526}
}

@article{BudishKessler2022,
  author  = {Budish, Eric and Kessler, Judd B.},
  title   = {Can Market Participants Report Their Preferences Accurately (Enough)?},
  journal = {Management Science},
  year    = {2022},
  volume  = {68},
  number  = {2},
  pages   = {1107--1130}
}

@article{ReesJones2018,
  author  = {Rees-Jones, Alex},
  title   = {Suboptimal Behavior in Strategy-Proof Mechanisms: Evidence from the Residency Match},
  journal = {Games and Economic Behavior},
  year    = {2018},
  volume  = {108},
  pages   = {317--330}
}

@article{ReesJonesShorrer2023,
  author  = {Rees-Jones, Alex and Shorrer, Ran I.},
  title   = {Behavioral Economics in Education Market Design: A Forward-Looking Review},
  journal = {Journal of Political Economy Microeconomics},
  year    = {2023},
  volume  = {1},
  number  = {3},
  pages   = {557--613}
}

@article{ShorrerSovago2023,
  author  = {Shorrer, Ran I. and S{\'o}v{\'a}g{\'o}, S{\'a}ndor},
  title   = {Dominated Choices in a Strategically Simple College Admissions Environment},
  journal = {Journal of Political Economy Microeconomics},
  year    = {2023},
  volume  = {1},
  number  = {4},
  pages   = {781--807}
}

@article{DreyfussHeffetzRabin2022,
  author  = {Dreyfuss, Bnaya and Heffetz, Ori and Rabin, Matthew},
  title   = {Expectations-Based Loss Aversion May Help Explain Seemingly Dominated Choices in Strategy-Proof Mechanisms},
  journal = {American Economic Journal: Microeconomics},
  year    = {2022},
  volume  = {14},
  number  = {4},
  pages   = {515--555}
}

@article{HakimovKubler2021,
  author  = {Hakimov, Rustamdjan and K{\"u}bler, Dorothea},
  title   = {Experiments on Centralized School Choice and College Admissions: A Survey},
  journal = {Experimental Economics},
  year    = {2021},
  volume  = {24},
  pages   = {434--488}
}

@article{BoHakimov2024,
  author  = {B{\'o}, In{\'a}cio and Hakimov, Rustamdjan},
  title   = {Pick-an-Object Mechanisms},
  journal = {Management Science},
  year    = {2024},
  volume  = {70},
  number  = {7},
  pages   = {4693--4721}
}

@article{GrenetHeKubler2022,
  author  = {Grenet, Julien and He, YingHua and K{\"u}bler, Dorothea},
  title   = {Preference Discovery in University Admissions: The Case for Dynamic Multioffer Mechanisms},
  journal = {Journal of Political Economy},
  year    = {2022},
  volume  = {130},
  number  = {6},
  pages   = {1427--1476}
}

@article{Narita2018,
  title={Match or mismatch? Learning and inertia in school choice},
  author={Narita, Yusuke},
  journal={Mimeo},
  year={2018}
}

@article{MeisnerWangenheim2023,
  author  = {Meisner, Vincent and von Wangenheim, Jonas},
  title   = {Loss Aversion in Strategy-Proof School-Choice Mechanisms},
  journal = {Journal of Economic Theory},
  year    = {2023},
  volume  = {207},
  pages   = {105588}
}

@article{Gerasimou2018,
  author  = {Gerasimou, Georgios},
  title   = {Indecisiveness, Undesirability and Overload Revealed through Rational Choice Deferral},
  journal = {The Economic Journal},
  year    = {2018},
  volume  = {128},
  number  = {614},
  pages   = {2450--2479}
}

@article{AgranovOrtoleva2017,
  author  = {Agranov, Marina and Ortoleva, Pietro},
  title   = {Stochastic Choice and Preferences for Randomization},
  journal = {Journal of Political Economy},
  year    = {2017},
  volume  = {125},
  number  = {1},
  pages   = {40--68}
}

@article{TverskyShafir1992,
  author  = {Tversky, Amos and Shafir, Eldar},
  title   = {Choice under Conflict: The Dynamics of Deferred Decision},
  journal = {Psychological Science},
  year    = {1992},
  volume  = {3},
  number  = {6},
  pages   = {358--361}
}

@article{CaspariKhanna2024,
  author  = {Caspari, Gian and Khanna, Manshu},
  title   = {Nonstandard Choice in Matching Markets},
  journal={International Economic Review},
  volume={66},
  number={2},
  pages={757--786},
  year={2025},
  publisher={Wiley Online Library}
}

@article{pan2019instability,
  title={The instability of matching with overconfident agents},
  author={Pan, Siqi},
  journal={Games and Economic Behavior},
  volume={113},
  pages={396--415},
  year={2019},
  publisher={Elsevier}
}

@article{meisner2026confidence,
  title={Confidence and information in strategy-proof school choice},
  author={Meisner, Vincent and S{\"u}er, M{\"u}ge and Tolksdorf, Michel and Tominaj, Sokol},
  year={2026},
  publisher={Kiel, Hamburg: ZBW-Leibniz Information Centre for Economics}
}

@Article{bo2020iterative,
  author    = {B{\'o}, In{\'a}cio and Hakimov, Rustamdjan},
  title     = {Iterative versus standard deferred acceptance: Experimental evidence},
  journal   = {The Economic Journal},
  year      = {2020},
  volume    = {130},
  number    = {626},
  pages     = {356--392},
  publisher = {Oxford University Press},
}

@Article{meisner2023report,
  author    = {Meisner, Vincent},
  journal   = {Management Science},
  title     = {Report-dependent utility and strategy-proofness},
  year      = {2023},
  number    = {5},
  pages     = {2733--2745},
  volume    = {69},
  publisher = {INFORMS},
}

@article{chen2006school,
  title={School choice: an experimental study},
  author={Chen, Yan and S{\"o}nmez, Tayfun},
  journal={Journal of Economic Theory},
  volume={127},
  number={1},
  pages={202--231},
  year={2006},
  publisher={Elsevier}
}

@article{kloosterman2025rankings,
  title={Rankings-dependent preferences: A real goods matching experiment},
  author={Kloosterman, Andrew and Troyan, Peter},
  journal={Games and Economic Behavior},
  year={2025},
  publisher={Elsevier}
}

@article{chen2021information,
  title={Information acquisition and provision in school choice: an experimental study},
  author={Chen, Yan and He, Yinghua},
  journal={Journal of Economic Theory},
  volume={197},
  pages={105345},
  year={2021},
  publisher={Elsevier}
}

@article{hakimov2023costly,
  title={Costly information acquisition in centralized matching markets},
  author={Hakimov, Rustamdjan and K{\"u}bler, Dorothea and Pan, Siqi},
  journal={Quantitative Economics},
  volume={14},
  number={4},
  pages={1447--1490},
  year={2023},
  publisher={Wiley Online Library}
}

@article{hassidim2017mechanism,
  title={The mechanism is truthful, why aren't you?},
  author={Hassidim, Avinatan and Marciano, D{\'e}borah and Romm, Assaf and Shorrer, Ran I},
  journal={American Economic Review},
  volume={107},
  number={5},
  pages={220--224},
  year={2017},
  publisher={American Economic Association 2014 Broadway, Suite 305, Nashville, TN 37203}
}

@article{rees2018experimental,
  title={An experimental investigation of preference misrepresentation in the residency match},
  author={Rees-Jones, Alex and Skowronek, Samuel},
  journal={Proceedings of the National Academy of Sciences},
  volume={115},
  number={45},
  pages={11471--11476},
  year={2018},
  publisher={National Academy of Sciences}
}

@book{kahneman2011book,
  author    = {Daniel Kahneman},
  title     = {Thinking, Fast and Slow},
  publisher = {Farrar, Straus and Giroux},
  year      = {2011},
  address   = {New York, NY},
  isbn      = {978-0-374-53355-7}
}

@inproceedings{romano2006,
  author    = {Romano, Jeanine and Kromrey, Jeffrey D. and Coraggio, Jesse and Skowronek, Jeff},
  title     = {Appropriate statistics for ordinal level data: Should we really be using t-test and {C}ohen's d for evaluating group differences on the {NSSE} and other surveys?},
  booktitle = {Annual Meeting of the Florida Association of Institutional Research},
  address   = {Cocoa Beach, FL},
  year      = {2006}
}

@article{basteck2023aiding,
  title={Aiding applicants: Leveling the playing field within the immediate acceptance mechanism},
  author={Basteck, Christian and Mantovani, Marco},
  journal={Review of Economic Design},
  volume={27},
  number={1},
  pages={187--220},
  year={2023},
  publisher={Springer}
}

@article{basteck2018cognitive,
  title={Cognitive ability and games of school choice},
  author={Basteck, Christian and Mantovani, Marco},
  journal={Games and Economic Behavior},
  volume={109},
  pages={156--183},
  year={2018},
  publisher={Elsevier}
}

@article{pathak2008leveling,
  title={Leveling the playing field: Sincere and sophisticated players in the Boston mechanism},
  author={Pathak, Parag A and S{\"o}nmez, Tayfun},
  journal={American Economic Review},
  volume={98},
  number={4},
  pages={1636--1652},
  year={2008},
  publisher={American Economic Association}
}

@article{klijn2019static,
  title={Static versus dynamic deferred acceptance in school choice: Theory and experiment},
  author={Klijn, Flip and Pais, Joana and Vorsatz, Marc},
  journal={Games and Economic Behavior},
  volume={113},
  pages={147--163},
  year={2019},
  publisher={Elsevier}
}

\newpage
\appendix

\section{Qualitative Pre-Mechanism Survey}
\label{app:qualitative-survey}

This appendix documents the qualitative survey introduced in Section~\ref{sec:design}. The survey was administered after the incentivized practice rounds and the practice-stage feedback screen, and before participants saw any mechanism-stage instructions. It comprised 11 screens: eight covered the task-level attribute elicitation, two covered the attribute-importance elicitation, and one covered bundle-level variety preference. The elicited scores enter the participant-specific score ranking used to construct the score-fit and score-deviation outcomes, and the variety-preference item enters heterogeneity analyses as a participant-level covariate.

\subsection{Task-level attribute elicitation}

Participants first read an introduction screen defining the four attributes used throughout the survey:
\begin{itemize}
    \item \textbf{Cognitive Ease}: how effortless the task felt---how little mental effort, memory load, or switching between steps it required.
    \item \textbf{Engagement}: how enjoyable or stimulating the task felt---whether it was fun or kept the participant's interest.
    \item \textbf{Confidence}: how sure the participant felt that they could do the task well.
    \item \textbf{Time Efficiency}: how quick and smooth the task felt and how long it took.
\end{itemize}

For each of the four attributes, the participant then completed two consecutive screens. The first was a ranking screen: the four tasks (math-memory, spot-the-difference, quiz, emotion-recognition) appeared as draggable items on the left and were placed into ranks~1 through~4 on the right. Per-attribute polarity was made explicit on screen; for instance, the Cognitive Ease ranking instructed participants to ``place the task that is least mentally exhausting at the top'' (Rank~1), while the Engagement ranking instructed them to ``place the task that you find most engaging at the top.'' The second screen elicited a cardinal score from 1 to 10 for each of the four tasks on the same attribute. The instruction emphasized absolute rather than comparative judgment: ``These scores should reflect your personal view of how strongly each task expresses this attribute.'' The four attributes were elicited in the fixed order Cognitive Ease, Engagement, Confidence, Time Efficiency.

After the eight task-level pages, participants ranked the four attributes themselves by their importance with Rank~1 the most important and Rank~4 the least important. The subsequent page asked participants to assign each attribute an importance score from 1 to 10. 

The eleventh and final screen of the survey presented a single 5-point Likert item capturing the participant's preference between bundles containing different tasks (``Mixed-Task bundle'') and bundles containing repeats of the same task (``Same-Task bundle''), with worked example illustrations of each. The five response options were:
\begin{enumerate}
    \item I strongly prefer a bundle with different tasks,
    \item I mildly prefer a bundle with different tasks,
    \item I am indifferent,
    \item I mildly prefer a bundle with the same tasks,
    \item I strongly prefer a bundle with the same tasks.
\end{enumerate}
Variety is a bundle-level property and is therefore not well-defined at the task level; consequently it was elicited as a stand-alone item rather than as a fifth attribute in the task-level instrument.

\subsection{Use of the elicited scores in the analysis}\label{Appendix_Algorithm}

The task-by-attribute score matrix and the attribute-importance scores together yield a participant-specific scalar score for each candidate bundle in each menu. Formally, let $c_a \in \{1,\dots,10\}$ denote the participant's importance score for attribute $a \in \{\mathrm{CE}, \mathrm{Eng}, \mathrm{Conf}, \mathrm{TE}\}$ and let $s_{t,a} \in \{1,\dots,10\}$ denote their task-by-attribute score for task $t$ on attribute $a$. The normalized attribute weight is $w_a = c_a / \sum_{a'} c_{a'}$, so that $\sum_a w_a = 1$. The participant-specific task score is the weighted sum of the four task-by-attribute scores,
\[
S_t \;=\; \sum_{a} w_a \, s_{t,a},
\]
and the participant-specific score of a bundle $b$ consisting of tasks $\{t_1,\dots,t_k\}$ is the sum of its component task scores, modulated by a variety multiplier $v(b)$:
\[
S_b \;=\; v(b) \cdot \sum_{i=1}^{k} S_{t_i}.
\]
The variety multiplier $v(b)$ equals 1 for single-task bundles ($k=1$). For multi-task bundles ($k>1$), let $\gamma \in \{0.8, 0.9, 1.0, 1.1, 1.2\}$ denote the participant's stated variety preference, with $\gamma=1.2$ corresponding to ``I strongly prefer a bundle with different tasks'' and $\gamma=0.8$ to ``I strongly prefer a bundle with the same tasks.'' Then $v(b) = \gamma$ if the component tasks in $b$ are of different types, and $v(b) = 2 - \gamma$ if the component tasks are repeats of the same task type. The encoding is symmetric: a participant with $\gamma=1.2$ has different-task bundles boosted by 20\% and same-task bundles penalized by 20\%, and vice versa for $\gamma=0.8$. Bundle scores in turn determine the participant's stated-preference ranking of the bundles in each menu, which enters the preregistered score-fit and score-deviation outcomes (see Section~\ref{sec:hypotheses}). Because the difficulty level of a task does not enter $S_b$, bundles that differ only in difficulty receive identical scores; Appendix Table~\ref{tab:errortypes-tiebreak} reports the score-deviation results with such ties broken in favor of the easier, dominating bundle. The bundle-level variety preference enters as a participant-level covariate in the preference-type heterogeneity analysis.

\newpage
\section{Robustness}
\label{app:robustness}

Five robustness checks support the main conclusions. First, the ability result does not depend on how the imbalance documented in Section~\ref{sec:design} is handled (Table~\ref{tab:sens}). The estimated harm to low-ability participants is $+0.063$ ($p=0.001$) in the sample as reported; $+0.071$ ($p<0.001$) controlling for the continuous ability index within the stratum; $+0.056$ ($p=0.004$) when ability is split within rather than across arms; and $+0.053$ ($p=0.008$) when the two sessions containing the anomalous learning-stage records are excluded, with the corresponding interaction moving from $-0.744$ to $-0.589$ but remaining significant ($p=0.024$). The twenty-two participants involved are retained in all reported estimates; the exclusion is reported only as a sensitivity check, since no technical fault behind their records could be established. Second, the mechanism-by-menu-size interaction is small and statistically insignificant (Table~\ref{tab:mechxmenu}), so the ability result is not merely driven by the number of binary comparisons embedded in \BSD. Third, mechanism-by-gender interactions are also insignificant (Table~\ref{tab:gender}). Fourth, task-specific ability measures reproduce the qualitative pattern of the aggregate ability index, with the low-ability treatment difference most pronounced in the quiz task (Table~\ref{tab:task-ability}); the heterogeneity result is therefore not an artifact of how the ability index is constructed. Fifth, decision times are about five seconds longer per round under \BSD\ than under \SSD, 31.0 versus 26.4 seconds, with the gap present in both ability strata (Table~\ref{tab:time-combined}). This rules out a ``\BSD\ takes less time'' explanation for the choice-quality results; the additional deliberation does not buy low-ability participants more reliable choices.

\begin{table}[htbp]
\centering
\captionsetup{width=\linewidth, font=footnotesize}
\caption{Sensitivity of the ability result to the pre-treatment ability imbalance}
\label{tab:sens}
\footnotesize
\begin{tabular}{lccccc}
\toprule
 & Binary SD effect & (SE) & $p$ & Interaction & $N$ \\
\cmidrule(lr){2-4} \cmidrule(lr){5-5} \cmidrule(lr){6-6}
\midrule
As reported (pooled median split) & \num{0.063}\sym{***} & (\num{0.019}) & \num{0.001} & \num{-0.744}\sym{***} & \num{247} \\
Controlling for ability within stratum & \num{0.071}\sym{***} & (\num{0.020}) & \num{< 0.001} & --- & \num{247} \\
Median split taken within arm & \num{0.056}\sym{***} & (\num{0.019}) & \num{0.004} & --- & \num{248} \\
Excluding anomalous-record sessions & \num{0.053}\sym{***} & (\num{0.020}) & \num{0.008} & \num{-0.589}\sym{**} & \num{226} \\
\bottomrule
\end{tabular}
\captionsetup{width=\linewidth, font=scriptsize}
\caption*{\textit{Notes:} Each row reports the Binary SD effect on the participant-level flip rate among low-ability participants, under a different treatment of the pre-treatment ability imbalance documented in Section~\ref{sec:design}. Heteroskedasticity-robust (HC1) standard errors in parentheses. The interaction column reports the $\text{Binary SD} \times \text{Ability}$ coefficient from the choice-level mixed logit with random intercepts for participant and session arm, in the native units of the ability index, so that row~1 reproduces the interaction reported in Table~\ref{tab:ame}; it is reported only for the specifications in which it is defined. Row~1 is the specification reported in the main text. Row~2 adds the continuous ability index as a control inside the low-ability stratum. Row~3 takes the median split within each arm rather than pooling, which removes the compositional difference between arms by construction. Row~4 excludes the two sessions containing the anomalous learning-stage records; those participants are retained in every estimate reported elsewhere in the paper. $^{*}$ $p<0.10$, $^{**}$ $p<0.05$, $^{***}$ $p<0.01$.}
\end{table}

\begin{table}[htbp]
\centering
\captionsetup{width=\linewidth, font=footnotesize}
\caption{Decision time by mechanism, overall and by ability group (participant-level means, winsorized)}
\label{tab:time-combined}
\footnotesize
\begin{tabular}{lcccccccc}
\toprule
& Menu SD & (SD) & Binary SD & (SD) & $p$ (MWU) & Cliff's $\delta$ & $N$ (Menu SD) & $N$ (Binary SD) \\
\midrule
Overall & \num{26.4} & (\num{6.8}) & \num{31.0} & (\num{7.2}) & \num{< 0.001}*** & \num{-0.350} & \num{241} & \num{253} \\
Low ability & \num{27.1} & (\num{6.9}) & \num{30.8} & (\num{7.5}) & \num{< 0.001}*** & \num{-0.284} & \num{131} & \num{116} \\
High ability & \num{25.6} & (\num{6.7}) & \num{31.3} & (\num{7.0}) & \num{< 0.001}*** & \num{-0.425} & \num{110} & \num{137} \\
\bottomrule
\end{tabular}
\captionsetup{width=\linewidth, font=scriptsize}
\caption*{\textit{Notes:} Standard deviations in parentheses. Means are participant-level averages of winsorized round-level decision times; the Overall row pools the participants in the ability sample, so $N_{\text{Overall}} = N_{\text{Low}} + N_{\text{High}}$. $p$-values from Mann--Whitney tests on participant-level means. $^{*}$ $p<0.10$, $^{**}$ $p<0.05$, $^{***}$ $p<0.01$.}
\end{table}

\section{Additional Tables and Figures}
\label{app:tables}

\begin{table}[htbp]
\centering
\captionsetup{width=\linewidth, font=footnotesize}
\caption{Conditional flip rates P($F \mid D, \text{off\_dom}$): dominance violation $\times$ dominator-at-revisit (pooled)}
\label{tab:flip_2x2}
\footnotesize
\begin{tabular}{lcc}
\toprule
 & \multicolumn{2}{c}{Dominator at revisit} \\
\cmidrule(lr){2-3}
 & off\_dom $= 0$ & off\_dom $= 1$ \\
\midrule
$D=0$ \quad P(F)  & \num{0.134} & --- \\
 \hspace*{1em} $N$ & (\num{6014}) &  \\
$D=1$ \quad P(F)  & \num{0.245} & \num{0.517} \\
 \hspace*{1em} $N$ & (\num{785}) & (\num{611}) \\
\bottomrule
\end{tabular}
\captionsetup{width=\linewidth, font=scriptsize}
\caption*{\textit{Notes:} Sample: revisit-eligible rounds ($N=7410$), pooled across mechanisms. Rows partition by whether the initial choice violates dominance ($D$); columns partition by whether the bundle shown at revisit strictly dominates the initial choice (off\_dom). The $(D=0, \text{off\_dom}=1)$ cell is empty by construction: if the initial choice was not dominated, the menu contains no bundle that dominates it, and the offered alternative cannot be a dominator. Cell entries are participant-level flip rates Pr($F \mid D, \text{off\_dom}$) with cell counts $N$ in parentheses below. See Appendix Table~\ref{tab:flip_2x2_by_mech} for the by-mechanism decomposition. $^{*}$ $p<0.10$, $^{**}$ $p<0.05$, $^{***}$ $p<0.01$.}
\end{table}

% --- Hypothesis-test tables (referenced by main-text table footers) ---
\begin{table}[htbp]
\centering
\captionsetup{width=\linewidth, font=footnotesize}
\caption{Primary hypothesis test results (Holm-corrected)}
\label{tab:primary}
\footnotesize
\begin{tabular}{lccc}
\toprule
Hypothesis & Family & $p$ (raw) & $p$ (Holm) \\
\midrule
H1a: Flip rate & F1: Decision quality & \num{0.589} & \num{1.000} \\
H1b: Dominance violations & F1: Decision quality & \num{0.555} & \num{1.000} \\
H2a: Complexity trend (Page's L) & F2: Complexity & \num{1.000} & \num{1.000} \\
H2b: Menu size association (Kendall's tau) & F2: Complexity & \num{< 0.001} & \num{< 0.001}*** \\
H3a: Expected profit & F3: Allocation outcomes & \num{0.043} & \num{0.173} \\
H3b: Realized payoff & F3: Allocation outcomes & \num{0.607} & \num{1.000} \\
H3c: Score fit (final bundle) & F3: Allocation outcomes & \num{0.157} & \num{0.470} \\
H3d: Market dispersion (Fligner-Killeen) & F3: Allocation outcomes & \num{0.607} & \num{1.000} \\
H4: Internal consistency (Spearman) & F4: Consistency (single test) & \num{0.780} & \num{0.780} \\
\bottomrule
\end{tabular}
\captionsetup{width=\linewidth, font=scriptsize}
\caption*{\textit{Notes:}  Holm correction within families. $^{*}$ $p<0.10$, $^{**}$ $p<0.05$, $^{***}$ $p<0.01$.}
\end{table}

\clearpage
\begin{landscape}
\begin{table}[htbp]
\centering
\captionsetup{width=\linewidth, font=footnotesize}
\caption{Secondary hypothesis test results (BH-corrected)}
\label{tab:secondary}
\footnotesize
\begin{tabular}{lccc}
\toprule
Hypothesis & Family & $p$ (raw) & $p$ (BH) \\
\midrule
S1: Binary SD mitigates complexity & SF2: Binary SD x Complexity (single test) & \num{0.417} & \num{0.417} \\
S2: Score deviation & SF1: Suboptimal choices & \num{0.440} & \num{0.555} \\
S3: Payoff deviation & SF1: Suboptimal choices & \num{0.497} & \num{0.555} \\
S4: Dominance violations & SF1: Suboptimal choices & \num{0.555} & \num{0.555} \\
S5: Ability heterogeneity & SF3: Heterogeneity & \num{< 0.001} & \num{< 0.001}*** \\
S6: Preference-type heterogeneity & SF3: Heterogeneity & \num{0.920} & \num{0.920} \\
S7: Confidence (Binary SD > Menu SD) & SF4: Subjective alignment & \num{0.005} & \num{0.015}** \\
S7: Happiness (Binary SD > Menu SD) & SF4: Subjective alignment & \num{0.627} & \num{0.627} \\
S9: Soft-hard alignment (min across 6 correlations) & SF4: Subjective alignment & \num{0.010} & \num{0.015}** \\
S8: Market welfare and equality & SF5: Market welfare (single test) & \num{0.586} & \num{0.586} \\
\bottomrule
\end{tabular}
\captionsetup{width=\linewidth, font=scriptsize}
\caption*{\textit{Notes:}  Benjamini--Hochberg correction within families. $^{*}$ $p<0.10$, $^{**}$ $p<0.05$, $^{***}$ $p<0.01$.}
\end{table}

\end{landscape}
\begin{table}[htbp]
\centering
\captionsetup{width=\linewidth, font=footnotesize}
\caption{Effect sizes: Cliff's delta (Binary SD vs.\ Menu SD)}
\label{tab:effects}
\footnotesize
\begin{tabular}{lcc}
\toprule
Outcome & Cliff's $\delta$ & Magnitude \\
\midrule
Flip rate (H1a) & \num{-0.028} & negligible \\
Dominance violations (H1b) & \num{0.030} & negligible \\
Expected profit (H3a) & \num{-0.105} & negligible \\
Realized payoff (H3b) & \num{0.023} & negligible \\
Score fit (H3c) & \num{-0.074} & negligible \\
\bottomrule
\end{tabular}
\end{table}

{\setlength{\tabcolsep}{3pt}%
\begin{table}[htbp]
\centering
\captionsetup{width=\linewidth, font=footnotesize}
\caption{Error-type decomposition of initial choices and reconsideration flip rates}
\label{tab:errortypes}
\footnotesize
\begin{tabular}{lcccccc}
\toprule
 & \multicolumn{2}{c}{Share of choices} & Menu SD flip rate & Binary SD flip rate & Binary SD$-$Menu SD & Holm $p$ \\
\cmidrule(lr){2-3} \cmidrule(lr){4-5} \cmidrule(lr){6-6} \cmidrule(lr){7-7}
 & Initial & Final & & & & \\
\midrule
No measured error & \num{0.472} & \num{0.538} & \num{0.082} & \num{0.095} & \num{0.012} & \num{0.996} \\
 & & & (\num{0.008}) & (\num{0.010}) & (\num{0.013}) & [\num{0.332}] \\
Pure dominance violation & \num{0.117} & \num{0.120} & \num{0.444} & \num{0.408} & \num{-0.037} & \num{0.996} \\
 & & & (\num{0.033}) & (\num{0.036}) & (\num{0.049}) & [\num{0.451}] \\
Pure score deviation & \num{0.339} & \num{0.285} & \num{0.189} & \num{0.204} & \num{0.015} & \num{0.996} \\
 & & & (\num{0.015}) & (\num{0.015}) & (\num{0.021}) & [\num{0.489}] \\
Compound error & \num{0.071} & \num{0.057} & \num{0.205} & \num{0.319} & \num{0.114} & \num{0.041}\sym{**} \\
 & & & (\num{0.027}) & (\num{0.035}) & (\num{0.044}) & [\num{0.010}] \\
\bottomrule
\end{tabular}
\captionsetup{width=\linewidth, font=scriptsize}
\caption*{\textit{Notes:} Initial choices are classified by whether they are objectively dominated ($D$) and/or deviate from the participant's elicited score model ($M$). The \textit{Initial} share column gives the distribution of initial choices over the four types; the \textit{Final} column classifies, by the same rules and against the same menu, the bundle each round ends with after reconsideration (the offered alternative if the initial choice was reversed, the initial choice otherwise). Reconsideration flip rates are the share of initial choices later reversed. Standard errors in parentheses below cell means and below the Binary SD$-$Menu SD difference are participant-clustered (HC1) from a cell-level OLS regression of \texttt{flip} on the mechanism indicator. Uncorrected two-sided $p$-values are reported in brackets below the Holm column; asterisks reflect the Holm-adjusted $p$ across the four within-table Binary SD$-$Menu SD tests. Marginal conditional probabilities: $\Pr(F \mid D)=\num{0.364}$; $\Pr(F \mid M, \lnot D)=\num{0.197}$; $\Pr(F \mid \lnot D, \lnot M)=\num{0.089}$. $^{*}$ $p<0.10$, $^{**}$ $p<0.05$, $^{***}$ $p<0.01$.}
\end{table}
}
\begin{table}[htbp]
\centering
\captionsetup{width=\linewidth, font=footnotesize}
\caption{Error-type decomposition with ties in the score model broken toward the easier bundle}
\label{tab:errortypes-tiebreak}
\footnotesize
\begin{tabular}{lcccccc}
\toprule
 & \multicolumn{2}{c}{Share of choices} & Menu SD flip rate & Binary SD flip rate & Binary SD$-$Menu SD & Holm $p$ \\
\cmidrule(lr){2-3} \cmidrule(lr){4-5} \cmidrule(lr){6-6} \cmidrule(lr){7-7}
 & Initial & Final & & & & \\
\midrule
No measured error & \num{0.472} & \num{0.538} & \num{0.082} & \num{0.095} & \num{0.012} & \num{0.996} \\
 & & & (\num{0.008}) & (\num{0.010}) & (\num{0.013}) & [\num{0.332}] \\
Pure score deviation & \num{0.339} & \num{0.285} & \num{0.189} & \num{0.204} & \num{0.015} & \num{0.996} \\
 & & & (\num{0.015}) & (\num{0.015}) & (\num{0.021}) & [\num{0.489}] \\
Compound error & \num{0.188} & \num{0.177} & \num{0.352} & \num{0.375} & \num{0.023} & \num{0.996} \\
 & & & (\num{0.025}) & (\num{0.026}) & (\num{0.036}) & [\num{0.533}] \\
\bottomrule
\end{tabular}
\captionsetup{width=\linewidth, font=scriptsize}
\caption*{\textit{Notes:} Replicates Table~\ref{tab:errortypes}, including the initial and final share columns, with a score-deviation indicator $M^{\varepsilon}$ that breaks ties in the elicited score in favor of the easier bundle. The score depends only on task composition, so bundles that differ only in difficulty (e.g.\ Math~1 and Math~2) tie exactly; \num{59.6}\% of menus have a tied top score. $M^{\varepsilon}=1$ if the chosen bundle is not top-scoring, or is top-scoring but component-wise dominated by another top-scoring bundle (equivalently, an arbitrarily small bonus is added to the score of a bundle that dominates another). This reclassifies \num{11.7}\% of all choices as deviations. Because a dominating bundle always has the same score as the bundle it dominates, $D=1$ implies $M^{\varepsilon}=1$ and the pure-dominance cell is empty by construction. Standard errors in parentheses are participant-clustered (HC1); uncorrected two-sided $p$-values in brackets; asterisks reflect the Holm-adjusted $p$ across the 3 within-table Binary SD$-$Menu SD tests. Marginal conditional probabilities: $\Pr(F \mid D)=\num{0.364}$; $\Pr(F \mid M^{\varepsilon}, \lnot D)=\num{0.197}$; $\Pr(F \mid \lnot D, \lnot M^{\varepsilon})=\num{0.089}$. $^{*}$ $p<0.10$, $^{**}$ $p<0.05$, $^{***}$ $p<0.01$.}
\end{table}

\begin{table}[htbp]
\centering
\captionsetup{width=\linewidth, font=footnotesize}
\caption{Market-level summary statistics}
\label{tab:market}
\footnotesize
\begin{tabular}{lcccc}
\toprule
& Menu SD & Binary SD & $p$ (MWU) & $p$ (Holm) \\
\midrule
Welfare (EUR) & \num{10.545} & \num{9.455} & \num{0.586} & \num{0.594} \\
& (\num{5.733}) & (\num{3.698}) & & \\
Dispersion (EUR) & \num{8.415} & \num{10.406} & \num{0.297} & \num{0.594} \\
& (\num{4.256}) & (\num{0.939}) & & \\
$N$ (markets) & \num{11} & \num{11} & & \\
\bottomrule
\end{tabular}
\captionsetup{width=\linewidth, font=scriptsize}
\caption*{\textit{Notes:} Markets are the live five-person markets recovered from the recorded menus: one per session and mechanism in the 11 sessions whose market structure is recoverable (sessions up to 15 January 2026). Welfare is the mean realized payoff per participant within a market; dispersion is the within-market standard deviation of realized payoffs. Both quantities are in EUR per participant. Standard deviations across markets in parentheses. Mann--Whitney $U$ tests of Menu SD vs.\ Binary SD at the market level for each outcome. $p$ (Holm) applies a Holm correction to the two tests in this table; asterisks reflect the Holm-corrected $p$-value. $^{*}$ $p<0.10$, $^{**}$ $p<0.05$, $^{***}$ $p<0.01$.}
\end{table}

%\input{Output/Tables/T7_H1_logit_models}

% --- Extension/robustness tables ---
\begin{table}[htbp]
\centering
\captionsetup{width=\linewidth, font=footnotesize}
\caption{Outcomes by complexity level and mechanism}
\label{tab:complexity}
\footnotesize
\begin{tabular}{lcccc}
\toprule
& \multicolumn{2}{c}{Menu SD}& \multicolumn{2}{c}{Binary SD}\\
Complexity & Flip rate & Dom.\ viol. & Flip rate & Dom.\ viol. \\
\midrule
Low & \num{0.230} & \num{0.198} & \num{0.206} & \num{0.193} \\
& (\num{0.421}) & (\num{0.399}) & (\num{0.405}) & (\num{0.395}) \\
Medium & \num{0.134} & \num{0.181} & \num{0.179} & \num{0.195} \\
& (\num{0.341}) & (\num{0.385}) & (\num{0.383}) & (\num{0.397}) \\
High & \num{0.147} & \num{0.188} & \num{0.167} & \num{0.175} \\
& (\num{0.354}) & (\num{0.391}) & (\num{0.373}) & (\num{0.380}) \\
\bottomrule
\end{tabular}
\captionsetup{width=\linewidth, font=scriptsize}
\caption*{\textit{Notes:} Standard deviations in parentheses. $^{*}$ $p<0.10$, $^{**}$ $p<0.05$, $^{***}$ $p<0.01$.}
\end{table}

\begin{table}[htbp]
\centering
\captionsetup{width=\linewidth, font=footnotesize}
\caption{Complexity $\times$ menu size logit regression}
\label{tab:h2reg}
\footnotesize
\begin{tabular}{lcc}
\toprule
& Coefficient & Odds ratio \\
\midrule
Intercept & \num{-2.362}*** & \num{0.094} \\
& (\num{0.200}) & \\
Medium & \num{0.331} & \num{1.392} \\
& (\num{0.286}) & \\
High & \num{0.331} & \num{1.392} \\
& (\num{0.286}) & \\
Menu size & \num{0.178}*** & \num{1.195} \\
& (\num{0.036}) & \\
Medium $\times$ Menu size & \num{-0.152}*** & \num{0.859} \\
& (\num{0.054}) & \\
High $\times$ Menu size & \num{-0.152}*** & \num{0.859} \\
& (\num{0.054}) & \\
Num.\ obs. & \num{7410} & \\
\bottomrule
\end{tabular}
\captionsetup{width=\linewidth, font=scriptsize}
\caption*{\textit{Notes:}  Standard errors in parentheses. Random intercepts for participant and session arm. $^{*}$ $p<0.10$, $^{**}$ $p<0.05$, $^{***}$ $p<0.01$.}
\end{table}

\begin{table}[htbp]
\centering
\captionsetup{width=\linewidth, font=footnotesize}
\caption{Binary SD $\times$ complexity interaction regression}
\label{tab:s1reg}
\footnotesize
\begin{tabular}{lcc}
\toprule
& Coefficient & Odds ratio \\
\midrule
Intercept & \num{-1.707}*** & \num{0.181} \\
& (\num{0.189}) & \\
Binary SD & \num{-0.344} & \num{0.709} \\
& (\num{0.262}) & \\
Medium & \num{-0.713}*** & \num{0.490} \\
& (\num{0.112}) & \\
High & \num{-0.602}*** & \num{0.548} \\
& (\num{0.110}) & \\
Menu size & \num{0.066}** & \num{1.068} \\
& (\num{0.032}) & \\
Binary SD $\times$ Medium & \num{0.515}*** & \num{1.673} \\
& (\num{0.153}) & \\
Binary SD $\times$ High & \num{0.312}** & \num{1.367} \\
& (\num{0.153}) & \\
Binary SD $\times$ Menu size & \num{0.036} & \num{1.037} \\
& (\num{0.045}) & \\
Num.\ obs. & \num{7410} & \\
\bottomrule
\end{tabular}
\captionsetup{width=\linewidth, font=scriptsize}
\caption*{\textit{Notes:}  Standard errors in parentheses. Random intercepts for participant and session arm. $^{*}$ $p<0.10$, $^{**}$ $p<0.05$, $^{***}$ $p<0.01$.}
\end{table}

\begin{table}[htbp]
\centering
\captionsetup{width=\linewidth, font=footnotesize}
\caption{Rank-specific outcomes by mechanism}
\label{tab:rank}
\footnotesize
\begin{tabular}{lcccccc}
\toprule
Outcome & Mechanism & Rank 1 & Rank 2 & Rank 3 & Rank 4 & Rank 5 \\
\midrule
\multirow{2}{*}{Expected payoff (EUR)} & Menu SD & \num{9.880} & \num{9.391} & \num{8.507} & \num{8.202} & \num{7.733} \\
& Binary SD & \num{10.395} & \num{9.961} & \num{8.901} & \num{9.280} & \num{8.522} \\
\multirow{2}{*}{Flip rate} & Menu SD & \num{0.170} & \num{0.194} & \num{0.173} & \num{0.181} & \num{0.134} \\
& Binary SD & \num{0.207} & \num{0.177} & \num{0.217} & \num{0.186} & \num{0.133} \\
\multirow{2}{*}{Score deviation} & Menu SD & \num{0.400} & \num{0.447} & \num{0.472} & \num{0.405} & \num{0.364} \\
& Binary SD & \num{0.404} & \num{0.437} & \num{0.456} & \num{0.386} & \num{0.333} \\
\multirow{2}{*}{Dominance violations} & Menu SD & \num{0.344} & \num{0.232} & \num{0.172} & \num{0.111} & \num{0.087} \\
& Binary SD & \num{0.340} & \num{0.208} & \num{0.182} & \num{0.146} & \num{0.062} \\
\bottomrule
\end{tabular}
\end{table}

\begin{table}[htbp]
\centering
\captionsetup{width=\linewidth, font=footnotesize}
\caption{Flip rate by mechanism and menu size}
\label{tab:mechxmenu}
\footnotesize
\begin{tabular}{lcc}
\toprule
Menu size & Menu SD & Binary SD \\
\midrule
\num{3} & \num{0.134} & \num{0.133} \\
\num{4} & \num{0.181} & \num{0.186} \\
\num{5} & \num{0.173} & \num{0.217} \\
\num{6} & \num{0.194} & \num{0.177} \\
\num{7} & \num{0.170} & \num{0.207} \\
\bottomrule
\end{tabular}
\end{table}

\begin{table}[htbp]
\centering
\captionsetup{width=\linewidth, font=footnotesize}
\caption{Flip rate by gender and mechanism}
\label{tab:gender}
\footnotesize
\begin{tabular}{lcccc}
\toprule
Gender & Menu SD & Binary SD & $p$ (MWU) & Cliff's $\delta$ \\
\midrule
Female & \num{0.170} & \num{0.170} & \num{0.837} & \num{0.014} \\
& (\num{0.132}) & (\num{0.138}) & & \\
Male & \num{0.171} & \num{0.207} & \num{0.286} & \num{-0.089} \\
& (\num{0.145}) & (\num{0.180}) & & \\
\bottomrule
\end{tabular}
\captionsetup{width=\linewidth, font=scriptsize}
\caption*{\textit{Notes:}  Standard deviations in parentheses. $^{*}$ $p<0.10$, $^{**}$ $p<0.05$, $^{***}$ $p<0.01$.}
\end{table}

\begin{table}[htbp]
\centering
\captionsetup{width=\linewidth, font=footnotesize}
\caption{Outcomes by ability group and mechanism, with stratified tests}
\label{tab:ability-combined}
\footnotesize

\begin{tabular}{lcccc}
\multicolumn{5}{l}{\textit{Panel A. Low ability}} \\
\toprule
Outcome & Menu SD & Binary SD & $p$ (MWU) & Cliff's $\delta$ \\
\midrule
Flip rate & \num{0.172} & \num{0.235} & \num{0.006}*** & \num{-0.202} \\
Score deviation & \num{0.423} & \num{0.409} & \num{0.554} & \num{0.044} \\
Dom.\ violations & \num{0.195} & \num{0.246} & \num{0.032}** & \num{-0.157} \\
Expected profit & \num{6.737} & \num{6.674} & \num{0.258} & \num{0.084} \\
Realized payoff & \num{8.855} & \num{8.966} & \num{0.931} & \num{-0.006} \\
\bottomrule
\end{tabular}

\medskip

\begin{tabular}{lcccc}
\multicolumn{5}{l}{\textit{Panel B. High ability}} \\
\toprule
Outcome & Menu SD & Binary SD & $p$ (MWU) & Cliff's $\delta$ \\
\midrule
Flip rate & \num{0.168} & \num{0.141} & \num{0.167} & \num{0.101} \\
Score deviation & \num{0.410} & \num{0.399} & \num{0.653} & \num{0.033} \\
Dom.\ violations & \num{0.182} & \num{0.138} & \num{0.013}** & \num{0.179} \\
Expected profit & \num{11.131} & \num{11.730} & \num{0.043}** & \num{-0.150} \\
Realized payoff & \num{14.364} & \num{12.555} & \num{0.135} & \num{0.090} \\
\bottomrule
\end{tabular}

\captionsetup{width=\linewidth, font=scriptsize}
\caption*{\textit{Notes:} Participant-level means by ability stratum and mechanism for the five preregistered outcomes. $p$-values from within-stratum Mann--Whitney $U$ tests on participant-level means; Cliff's $\delta$ reports the rank-based effect size. The continuous-ability interaction tests are reported in Table~\ref{tab:ame}. $^{*}$ $p<0.10$, $^{**}$ $p<0.05$, $^{***}$ $p<0.01$.}
\end{table}

\begin{table}[htbp]
\centering
\captionsetup{width=\linewidth, font=footnotesize}
\caption{Task-specific ability splits}
\label{tab:task-ability}
\footnotesize
\begin{tabular}{lcccccc}
\toprule
& \multicolumn{3}{c}{Low ability}& \multicolumn{3}{c}{High ability}\\
Task & Binary SD & Menu SD & $p$ & Binary SD & Menu SD & $p$ \\
\midrule
Quiz & \num{0.221} & \num{0.169} & \num{0.018}** & \num{0.150} & \num{0.172} & \num{0.165} \\
Math & \num{0.220} & \num{0.183} & \num{0.123} & \num{0.148} & \num{0.153} & \num{0.751} \\
Emotion & \num{0.206} & \num{0.168} & \num{0.141} & \num{0.164} & \num{0.173} & \num{0.482} \\
Spot & \num{0.208} & \num{0.180} & \num{0.404} & \num{0.162} & \num{0.158} & \num{0.730} \\
\bottomrule
\end{tabular}
\captionsetup{width=\linewidth, font=scriptsize}
\caption*{\textit{Notes:}  Mann--Whitney $p$-values. $^{*}$ $p<0.10$, $^{**}$ $p<0.05$, $^{***}$ $p<0.01$.}
\end{table}

\begin{table}[htbp]
\centering
\captionsetup{width=\linewidth, font=footnotesize}
\caption{Conditional probability of being below-median on each task}
\label{tab:transition}
\footnotesize
\begin{tabular}{lcccc}
\toprule
Given low on & Quiz & Math & Emotion & Spot \\
\midrule
Quiz & \num{1.00} & \num{0.65} & \num{0.56} & \num{0.55} \\
Math & \num{0.61} & \num{1.00} & \num{0.55} & \num{0.59} \\
Emotion & \num{0.54} & \num{0.57} & \num{1.00} & \num{0.54} \\
Spot & \num{0.55} & \num{0.62} & \num{0.55} & \num{1.00} \\
\bottomrule
\end{tabular}
\end{table}

\begin{table}[htbp]
\centering
\captionsetup{width=\linewidth, font=footnotesize}
\caption{Mechanism $\times$ continuous-ability interaction across participant-level outcomes}
\label{tab:e8-continuous-ability}
\footnotesize
\begin{tabular}{l S[table-format=-1.4] S[table-format=1.4] S[table-format=1.4] l}
\toprule
\textbf{Model} & {\textbf{Estimate}} & {\textbf{SE}} & {\textbf{p-value}} & \textbf{Sig.} \\
\midrule
Flip rate (OLS) & \num{-0.1082} & \num{0.0221} & \num{0.0000} & *** \\
Dominance viol. (OLS) & \num{-0.0825} & \num{0.0270} & \num{0.0023} & *** \\
Score deviation (OLS) & \num{-0.0263} & \num{0.0357} & \num{0.4613} &  \\
Expected profit (OLS) & \num{3.1096} & \num{0.2643} & \num{0.0000} & *** \\
Confidence (OLS) & \num{-0.6398} & \num{0.2025} & \num{0.0017} & *** \\
Happiness (OLS) & \num{-0.4785} & \num{0.2079} & \num{0.0218} & ** \\
\bottomrule
\end{tabular}
\captionsetup{width=\linewidth, font=scriptsize}
\caption*{\textit{Notes:} Each row reports the coefficient on the $\text{mechanism} \times \text{ability\_mean}$ interaction from a participant-level OLS regression of the listed outcome on mechanism, the continuous practice-stage ability index (\texttt{ability\_mean}), and their interaction. No control variables. Significance: $^{*}$ $p<0.10$, $^{**}$ $p<0.05$, $^{***}$ $p<0.01$.}
\end{table}

\begin{table}[htbp]
\centering
\captionsetup{width=\linewidth, font=footnotesize}
\caption{Normalized dominance-violation incidence: observed Pr($D=1$) versus the per-round menu-based exposure baseline}
\label{tab:ndi}
\footnotesize
\begin{tabular}{lccccc}
\toprule
 & $N$ & Baseline & Pr($D=1$) & Excess & $p$ \\
\cmidrule(lr){2-2} \cmidrule(lr){3-3} \cmidrule(lr){4-4} \cmidrule(lr){5-5} \cmidrule(lr){6-6}
\midrule
\multicolumn{6}{l}{\textbf{Panel A: Pooled}} \\
Overall & \num{7410} & \num{0.346} & \num{0.188} & \num{-0.158}\sym{***} & \num{0.000} \\
 & & & & (\num{0.008}) & \\
\midrule
\multicolumn{6}{l}{\textbf{Panel B: By bundle complexity (pooled across mechanisms)}} \\
Low & \num{2470} & \num{0.321} & \num{0.196} & \num{-0.126}\sym{***} & \num{0.000} \\
 & & & & (\num{0.009}) & \\
Medium & \num{2470} & \num{0.366} & \num{0.188} & \num{-0.178}\sym{***} & \num{0.000} \\
 & & & & (\num{0.010}) & \\
High & \num{2470} & \num{0.351} & \num{0.181} & \num{-0.170}\sym{***} & \num{0.000} \\
 & & & & (\num{0.010}) & \\
\midrule
\multicolumn{6}{l}{\textbf{Panel C: By mechanism $\times$ bundle complexity}} \\
Menu SD, Low & \num{1205} & \num{0.323} & \num{0.198} & \num{-0.124}\sym{***} & \num{0.000} \\
 & & & & (\num{0.013}) & \\
Menu SD, Medium & \num{1205} & \num{0.367} & \num{0.181} & \num{-0.186}\sym{***} & \num{0.000} \\
 & & & & (\num{0.015}) & \\
Menu SD, High & \num{1205} & \num{0.345} & \num{0.188} & \num{-0.156}\sym{***} & \num{0.000} \\
 & & & & (\num{0.014}) & \\
Binary SD, Low & \num{1265} & \num{0.320} & \num{0.193} & \num{-0.127}\sym{***} & \num{0.000} \\
 & & & & (\num{0.014}) & \\
Binary SD, Medium & \num{1265} & \num{0.366} & \num{0.195} & \num{-0.170}\sym{***} & \num{0.000} \\
 & & & & (\num{0.014}) & \\
Binary SD, High & \num{1265} & \num{0.357} & \num{0.175} & \num{-0.183}\sym{***} & \num{0.000} \\
 & & & & (\num{0.015}) & \\
\bottomrule
\end{tabular}
\captionsetup{width=\linewidth, font=scriptsize}
\caption*{\textit{Notes:} Sample: revisit-eligible rounds ($N=7410$). For each round, the exposure baseline is the share of bundles in the realized menu that are strictly dominated by some other menu bundle (component-wise, same task multiset, all levels $\leq$, at least one $<$). Excess $=$ observed Pr($D=1$) $-$ baseline. A negative excess means participants pick dominated bundles less often than a uniform draw from the same menu would. Participant-clustered HC1 standard errors in parentheses below the excess; raw two-sided $p$ in the last column. \textit{Regression footer.} $\text{dom\_viol}_{ij} \sim \text{Mech} \times \text{Complexity} + \text{offset}(b_{ij})$, clustered HC1: Binary SD$-$Menu SD $\hat{\beta}=\num{0.019}$ ($p=\num{0.455}$); Complexity (ordinal, Low$=1$) $\hat{\beta}=\num{-0.016}$ ($p=\num{0.047}$); Mech $\times$ Complexity $\hat{\beta}=\num{-0.012}$ ($p=\num{0.302}$). $^{*}$ $p<0.10$, $^{**}$ $p<0.05$, $^{***}$ $p<0.01$.}
\end{table}

\begin{table}[htbp]
\centering
\captionsetup{width=\linewidth, font=footnotesize}
\caption{Revisit-page response time, flippers vs.\ non-flippers (subset: dominated initial choice, dominator shown at revisit)}
\label{tab:nf_times}
\footnotesize
\begin{tabular}{lcccccc}
\toprule
 & \multicolumn{2}{c}{Flippers} & \multicolumn{2}{c}{Non-flippers} & & \\
 & $N$ & Mean (sec) & $N$ & Mean (sec) & $\hat{\beta}_{\text{non-flip}}$ & $p$ \\
\cmidrule(lr){2-3} \cmidrule(lr){4-5} \cmidrule(lr){6-6} \cmidrule(lr){7-7}
\midrule
\multicolumn{7}{l}{\textbf{Panel A: Pooled}} \\
Overall & \num{316} & \num{6.45} & \num{295} & \num{6.47} & \num{-0.007} & \num{0.870} \\
 & & (\num{3.67}) & & (\num{4.00}) & (\num{0.041}) & \\
\midrule
\multicolumn{7}{l}{\textbf{Panel B: By mechanism}} \\
Menu SD & \num{158} & \num{6.65} & \num{122} & \num{6.76} & \num{0.030} & \num{0.585} \\
 & & (\num{4.01}) & & (\num{3.79}) & (\num{0.054}) & \\
Binary SD & \num{158} & \num{6.25} & \num{173} & \num{6.26} & \num{-0.028} & \num{0.636} \\
 & & (\num{3.30}) & & (\num{4.14}) & (\num{0.060}) & \\
\midrule
\multicolumn{7}{l}{\textbf{Panel C: By bundle complexity}} \\
Low & \num{151} & \num{6.42} & \num{119} & \num{6.11} & \num{-0.060} & \num{0.359} \\
 & & (\num{4.35}) & & (\num{4.61}) & (\num{0.066}) & \\
Medium & \num{95} & \num{5.75} & \num{98} & \num{6.29} & \num{0.055} & \num{0.300} \\
 & & (\num{2.22}) & & (\num{3.28}) & (\num{0.053}) & \\
High & \num{70} & \num{7.46} & \num{78} & \num{7.24} & \num{-0.031} & \num{0.626} \\
 & & (\num{3.47}) & & (\num{3.76}) & (\num{0.064}) & \\
\midrule
\multicolumn{7}{l}{\textbf{Panel D: By ability}} \\
High & \num{147} & \num{6.03} & \num{131} & \num{6.57} & \num{0.040} & \num{0.569} \\
 & & (\num{3.17}) & & (\num{4.39}) & (\num{0.070}) & \\
Low & \num{169} & \num{6.81} & \num{164} & \num{6.38} & \num{-0.048} & \num{0.327} \\
 & & (\num{4.04}) & & (\num{3.67}) & (\num{0.049}) & \\
\bottomrule
\end{tabular}
\captionsetup{width=\linewidth, font=scriptsize}
\caption*{\textit{Notes:} Subset: rounds in which the initial choice violated dominance and the bundle shown at revisit weakly dominates it ($N=611$ rounds, 268 participants). Revisit-page durations are computed as the difference between consecutive oTree page-completion timestamps and winsorized at the 99th percentile ($24.0$ sec). Within-cell standard deviations in parentheses below the means. $\hat{\beta}_{\text{non-flip}}$ is the coefficient on the non-flip indicator from $\log(\text{rev\_time}+1) \sim \text{non\_flip}$, fit separately within each row's sample; participant-clustered HC1 standard errors in parentheses below the coefficient; raw two-sided $p$ in the last column. A positive coefficient indicates that non-flippers spend longer on the revisit page than flippers; a negative coefficient indicates the opposite. \textit{Interaction tests (footer).} non\_flip $\times$ Mechanism: $\hat{\beta}=\num{-0.058}$, $p=\num{0.473}$. non\_flip $\times$ Complexity (ordinal, Low$=1$): $\hat{\beta}=\num{0.027}$, $p=\num{0.534}$. non\_flip $\times$ Low ability: $\hat{\beta}=\num{-0.088}$, $p=\num{0.304}$. $^{*}$ $p<0.10$, $^{**}$ $p<0.05$, $^{***}$ $p<0.01$.}
\end{table}

\begin{table}[htbp]
\centering
\captionsetup{width=\linewidth, font=footnotesize}
\caption{Choice mistakes by comprehension-check outcome and mechanism}
\label{tab:comp_mistakes}
\footnotesize
\begin{tabular}{lcccccccc}
\toprule
 & & & \multicolumn{2}{c}{Dominance viol.} & \multicolumn{2}{c}{Reversals} & \multicolumn{2}{c}{Payoff-dominated} \\
\cmidrule(lr){4-5} \cmidrule(lr){6-7} \cmidrule(lr){8-9}
 & $N$ & Share of sample & Rate & Share of total & Rate & Share of total & Rate & Share of total \\
\midrule
\multicolumn{9}{l}{\textbf{Panel A: Menu SD}} \\
Passed first & \num{192} & \num{0.80} & \num{0.169} & \num{0.71} & \num{0.164} & \num{0.77} & \num{0.557} & \num{0.77} \\
Failed once & \num{31} & \num{0.13} & \num{0.267} & \num{0.18} & \num{0.194} & \num{0.15} & \num{0.675} & \num{0.15} \\
Failed twice & \num{18} & \num{0.07} & \num{0.267} & \num{0.11} & \num{0.196} & \num{0.09} & \num{0.619} & \num{0.08} \\
\midrule
\multicolumn{9}{l}{\textbf{Panel B: Binary SD}} \\
Passed first & \num{190} & \num{0.75} & \num{0.153} & \num{0.61} & \num{0.164} & \num{0.67} & \num{0.554} & \num{0.71} \\
Failed once & \num{40} & \num{0.16} & \num{0.282} & \num{0.24} & \num{0.238} & \num{0.20} & \num{0.650} & \num{0.18} \\
Failed twice & \num{23} & \num{0.09} & \num{0.310} & \num{0.15} & \num{0.258} & \num{0.13} & \num{0.728} & \num{0.11} \\
\bottomrule
\end{tabular}
\captionsetup{width=\linewidth, font=scriptsize}
\caption*{\textit{Notes:} Participants are grouped by the outcome of the three-question comprehension check administered after the mechanism instructions (two attempts; nobody was excluded). \textit{Rate} is the mean participant-level share of the 15 rounds with a dominance violation, a reversal at revisit, or a payoff-dominated choice; \textit{Share of total} is the group's share of all such rounds within the mechanism, to be compared with its share of the sample. Among first-time passers the Binary SD $-$ Menu SD difference in participant-level rates is \num{-0.016} for dominance violations ($p=\num{0.160}$), \num{-0.001} for reversals ($p=\num{0.770}$) and \num{-0.002} for payoff-dominated choices ($p=\num{0.901}$; Mann--Whitney). If the double-failers erred at the first-time passers' rate, the overall dominance-violation rate would be \num{0.178} instead of \num{0.188}. $^{*}$ $p<0.10$, $^{**}$ $p<0.05$, $^{***}$ $p<0.01$.}
\end{table}

\begin{table}[htbp]
\centering
\captionsetup{width=\linewidth, font=footnotesize}
\caption{Response times on reversed and non-reversed rounds (all rounds)}
\label{tab:flip_times}
\footnotesize
\begin{tabular}{lcccccccc}
\toprule
 & \multicolumn{2}{c}{$N$ rounds} & \multicolumn{3}{c}{Initial choice page (sec)} & \multicolumn{3}{c}{Revisit page (sec)} \\
\cmidrule(lr){2-3} \cmidrule(lr){4-6} \cmidrule(lr){7-9}
 & Non-flip & Flip & Non-flip & Flip & $\hat{\beta}_{\text{flip}}$ & Non-flip & Flip & $\hat{\beta}_{\text{flip}}$ \\
\midrule
Pooled & \num{6096} & \num{1314} & \num{28.6} & \num{29.6} & \num{0.032}\sym{**} & \num{6.5} & \num{6.8} & \num{0.059}\sym{***} \\
 & & & & & (\num{0.014}) & & & (\num{0.013}) \\
Menu SD & \num{2999} & \num{616} & \num{26.1} & \num{28.1} & \num{0.043}\sym{**} & \num{6.8} & \num{7.3} & \num{0.081}\sym{***} \\
 & & & & & (\num{0.022}) & & & (\num{0.020}) \\
Binary SD & \num{3097} & \num{698} & \num{31.1} & \num{31.0} & \num{0.010} & \num{6.2} & \num{6.4} & \num{0.040}\sym{**} \\
 & & & & & (\num{0.015}) & & & (\num{0.017}) \\
\bottomrule
\end{tabular}
\captionsetup{width=\linewidth, font=scriptsize}
\caption*{\textit{Notes:} All 7410 rounds of the complete sample (494 participants), split by whether the initial choice was reversed at revisit. Page durations are differences between consecutive oTree page-completion timestamps (group wait pages booked separately) and winsorized at the 99th percentile. Means in seconds. $\hat{\beta}_{\text{flip}}$ is the coefficient on the reversal indicator from a participant fixed-effects regression of $\log(\text{time})$ ($\log(\text{time}+1)$ for the revisit page) on the indicator, complexity and rank dummies, fit within each row's sample; participant-clustered HC1 standard errors in parentheses. A positive coefficient means reversed rounds took longer. Reversal $\times$ Binary SD interaction: initial page $\hat{\beta}=\num{-0.036}$ ($p=\num{0.191}$), revisit page $\hat{\beta}=\num{-0.039}$ ($p=\num{0.138}$). \textit{Participant level:} mean initial-page time by quartile of the individual reversal rate (lowest to highest) is 29.3, 28.2, 29.0, 28.7 s (Kruskal--Wallis $p=\num{0.830}$) and mean revisit-page time is 6.5, 6.5, 6.5, 6.7 s ($p=\num{0.529}$); Spearman correlations of the reversal rate with the two times are \num{-0.004} ($p=\num{0.932}$) and \num{0.040} ($p=\num{0.371}$). $^{*}$ $p<0.10$, $^{**}$ $p<0.05$, $^{***}$ $p<0.01$.}
\end{table}

\begin{table}[htbp]
\centering
\captionsetup{width=\linewidth, font=footnotesize}
\caption{Response times by error type of the initial choice}
\label{tab:et_times}
\footnotesize
\begin{tabular}{lcccccccc}
\toprule
 & & & \multicolumn{2}{c}{Initial choice page (sec)} & \multicolumn{4}{c}{Revisit page (sec)} \\
\cmidrule(lr){4-5} \cmidrule(lr){6-9}
Initial choice & $N$ & Flip rate & Mean & $\hat{\beta}$ & Non-flip & Flip & $\hat{\beta}$ & $\hat{\beta}\mid$flip \\
\midrule
No measured error & \num{3500} & \num{0.089} & \num{28.1} & ref. & \num{6.2} & \num{7.0} & ref. & ref. \\
Pure dominance violation & \num{870} & \num{0.425} & \num{29.5} & \num{-0.003} & \num{6.9} & \num{6.6} & \num{0.037}\sym{**} & \num{0.021} \\
 & & & & (\num{0.018}) & & & (\num{0.016}) & (\num{0.017}) \\
Pure score deviation & \num{2514} & \num{0.197} & \num{29.1} & \num{0.017} & \num{6.7} & \num{6.9} & \num{0.063}\sym{***} & \num{0.057}\sym{***} \\
 & & & & (\num{0.011}) & & & (\num{0.011}) & (\num{0.011}) \\
Compound error & \num{526} & \num{0.262} & \num{31.0} & \num{0.018} & \num{6.9} & \num{6.9} & \num{0.054}\sym{***} & \num{0.045}\sym{**} \\
 & & & & (\num{0.022}) & & & (\num{0.020}) & (\num{0.020}) \\
\bottomrule
\end{tabular}
\captionsetup{width=\linewidth, font=scriptsize}
\caption*{\textit{Notes:} All 7410 rounds of the complete sample, classified by the error type of the initial choice as in Table~\ref{tab:errortypes}. Page durations are differences between consecutive oTree page-completion timestamps (group wait pages booked separately), winsorized at the 99th percentile; means in seconds. Revisit-page means are shown separately for rounds that were and were not reversed. $\hat{\beta}$ is the coefficient on the error-type indicator (reference: no measured error) from a participant fixed-effects regression of $\log(\text{time})$ ($\log(\text{time}+1)$ for the revisit page) on the four error-type indicators, complexity and rank dummies; the last column adds the reversal indicator, so it compares rounds with the same reversal outcome. Participant-clustered HC1 standard errors in parentheses. A positive coefficient means the error type took longer than a round with no measured error. $^{*}$ $p<0.10$, $^{**}$ $p<0.05$, $^{***}$ $p<0.01$.}
\end{table}

\begin{table}[htbp]
\centering
\captionsetup{width=\linewidth, font=footnotesize}
\caption{Decision time by menu size and complexity block (within participant)}
\label{tab:size_time}
\footnotesize
\begin{tabular}{lcc}
\toprule
Dependent variable: & \multicolumn{2}{c}{$\log(\text{decision time})$} \\
\cmidrule(lr){2-3}
 & Menu SD & Binary SD \\
 & (1) & (2) \\
\midrule
Menu size 4 & \num{0.121}\sym{***} & \num{0.189}\sym{***} \\
 & (\num{0.020}) & (\num{0.017}) \\
\addlinespace
Menu size 5 & \num{0.230}\sym{***} & \num{0.329}\sym{***} \\
 & (\num{0.022}) & (\num{0.016}) \\
\addlinespace
Menu size 6 & \num{0.331}\sym{***} & \num{0.507}\sym{***} \\
 & (\num{0.022}) & (\num{0.018}) \\
\addlinespace
Menu size 7 & \num{0.119}\sym{***} & \num{0.574}\sym{***} \\
 & (\num{0.028}) & (\num{0.024}) \\
\addlinespace
Medium complexity & \num{-0.120}\sym{***} & \num{-0.004} \\
 & (\num{0.016}) & (\num{0.014}) \\
\addlinespace
High complexity & \num{-0.058}\sym{***} & \num{0.131}\sym{***} \\
 & (\num{0.016}) & (\num{0.016}) \\
\midrule
$N$ rounds & \num{3615} & \num{3795} \\
Participants & \num{241} & \num{253} \\
Wald $p$ (sizes 4 = 5 = 6 = 7) & \num{< 0.001} & \num{< 0.001} \\
\bottomrule
\end{tabular}
\captionsetup{width=\linewidth, font=scriptsize}
\caption*{\textit{Notes:} Participant fixed-effects regressions of the log of the initial-choice page duration (differences between consecutive oTree page-completion timestamps, group wait pages booked separately, winsorized at the 99th percentile) on menu-size dummies (reference: three bundles) and complexity-block dummies (reference: one-task block), fit separately by mechanism. Participant-clustered HC1 standard errors in parentheses. Mean seconds by menu size (3 to 7): Menu SD 22.0, 25.3, 27.8, 30.6, 26.5; Binary SD 22.2, 26.9, 30.5, 36.1, 39.5. Mean seconds by complexity block (one, two, three tasks): Menu SD 28.3, 24.6, 26.3; Binary SD 30.3, 29.3, 33.6. Three qualifications: menu size equals $8-$rank, so it is collinear with within-block position; under Binary SD a $k$-bundle menu requires $k-1$ clicks, so part of its gradient is mechanical; and the seven-bundle page is the only choice page in each block not preceded by a group wait page, which lowers its duration under Menu SD. $^{*}$ $p<0.10$, $^{**}$ $p<0.05$, $^{***}$ $p<0.01$.}
\end{table}

\begin{table}[htbp]
\centering
\captionsetup{width=\linewidth, font=footnotesize}
\caption{Reversal and the score distance of the offered alternative}
\label{tab:gain}
\footnotesize
\begin{tabular}{lccc}
\toprule
Dependent variable: & \multicolumn{3}{c}{Reversal at revisit (logit, log-odds)} \\
\cmidrule(lr){2-4}
 & All rounds & Offer does not dominate & All rounds \\
 & (1) & (2) & (3) \\
\midrule
Score gain & \num{0.205}\sym{***} & \num{0.281}\sym{***} & \num{0.265}\sym{***} \\
 & (\num{0.029}) & (\num{0.033}) & (\num{0.031}) \\
\addlinespace
Binary SD & \num{0.083} & \num{0.156} & \num{0.063} \\
 & (\num{0.108}) & (\num{0.129}) & (\num{0.115}) \\
\addlinespace
Score gain $\times$ Binary SD & \num{0.047} & \num{0.032} & \num{0.044} \\
 & (\num{0.040}) & (\num{0.045}) & (\num{0.043}) \\
\addlinespace
Offered bundle dominates choice &  &  & \num{2.092}\sym{***} \\
 &  &  & (\num{0.102}) \\
\midrule
$N$ rounds & \num{7410} & \num{6799} & \num{7410} \\
Participants & \num{494} & \num{494} & \num{494} \\
Mean flip rate & \num{0.177} & \num{0.147} & \num{0.177} \\
\bottomrule
\end{tabular}
\captionsetup{width=\linewidth, font=scriptsize}
\caption*{\textit{Notes:} Mixed logits with random intercepts for participant and session arm. \textit{Score gain} is the participant's elicited score of the offered alternative minus that of the initial choice (score points; the offered alternative is the highest-scoring other bundle in the menu, so the gain is positive when the initial choice was not top-scoring and zero or negative otherwise). Column (2) drops the 611 rounds in which the offered bundle component-wise dominates the initial choice, so the score gain there is not confounded with an objective improvement; column (3) keeps all rounds and controls for that indicator. Among rounds without a dominating offer, the reversal rate by tercile of the score gain is 0.089, 0.120, 0.232 (tercile ranges -9.8 to 0.0; 0.0 to 0.7; 0.7 to 9.6 points). Standard errors in parentheses. $^{*}$ $p<0.10$, $^{**}$ $p<0.05$, $^{***}$ $p<0.01$.}
\end{table}

\begin{table}[htbp]
\centering
\captionsetup{width=\linewidth, font=footnotesize}
\caption{Conditional flip rates Pr($F \mid D, \text{off\_dom}$) by mechanism}
\label{tab:flip_2x2_by_mech}
\footnotesize
\begin{tabular}{lcc}
\toprule
 & \multicolumn{2}{c}{Dominator at revisit} \\
\cmidrule(lr){2-3}
 & off\_dom $= 0$ & off\_dom $= 1$ \\
\midrule
\multicolumn{3}{l}{\textbf{Panel A: Menu SD}} \\
$D=0$ \quad Pr(F)  & \num{0.128} & --- \\
 \hspace*{1em} $N$ & (\num{2931}) &  \\
$D=1$ \quad Pr(F)  & \num{0.205} & \num{0.564} \\
 \hspace*{1em} $N$ & (\num{404}) & (\num{280}) \\
\midrule
\multicolumn{3}{l}{\textbf{Panel B: Binary SD}} \\
$D=0$ \quad Pr(F)  & \num{0.140} & --- \\
 \hspace*{1em} $N$ & (\num{3083}) &  \\
$D=1$ \quad Pr(F)  & \num{0.286} & \num{0.477} \\
 \hspace*{1em} $N$ & (\num{381}) & (\num{331}) \\
\bottomrule
\end{tabular}
\captionsetup{width=\linewidth, font=scriptsize}
\caption*{\textit{Notes:} Sample: revisit-eligible rounds ($N=7410$), split by mechanism. Rows partition by whether the initial choice violates dominance ($D$); columns partition by whether the bundle shown at revisit strictly dominates the initial choice (off\_dom). The $(D=0, \text{off\_dom}=1)$ cell is empty by construction. Cell entries are participant-level flip rates with cell counts $N$ in parentheses below. The pooled version appears as Table~\ref{tab:flip_2x2} in the main text. $^{*}$ $p<0.10$, $^{**}$ $p<0.05$, $^{***}$ $p<0.01$.}
\end{table}

\begin{table}[!htbp]
\centering
\caption{Behavioral hypotheses and empirical tests}
\label{tab:hypothesis_map}
\begin{tabularx}{\textwidth}{p{0.18\textwidth}X X}
\toprule
Hypothesis & Behavioral mechanism & Main empirical test \\
\midrule
Hypothesis~\ref{hyp:inconsistency}: Choice inconsistency & Choices may not survive reconsideration even without strategic incentives. & Flip and dominance-violation rates in the mechanism and reconsideration stages. \\
Hypothesis~\ref{hyp:mechanism}: Simplification versus fragmentation & Pairwise comparisons reduce local cognitive load, or create anchoring and path dependence. & Flip and dominance rates by mechanism; mechanism \(\times\) ability and mechanism \(\times\) ability \(\times\) complexity models. \\
Hypothesis~\ref{hyp:complexity}: Complexity & Larger menus and larger bundles increase cognitive uncertainty. & Flip and dominance rates by menu size and bundle complexity. \\
\midrule
Secondary: error separability & Dominance and score-fit errors reflect different behavioral frictions. & Correlation of individual error propensities; mechanism interactions by error type. \\
Secondary: dominance detection & Objective dominance mistakes are easier to correct than score-fit mistakes. & \(\Pr(F_{ir}=1\mid D_{ir}=1)\) versus \(\Pr(F_{ir}=1\mid M_{ir}=1, D_{ir}=0)\). \\
Secondary: confidence & Calibrated confidence predicts fewer reversals; local simplicity may raise confidence without improving choice quality. & Confidence by mechanism and ability; flips conditional on reported confidence. \\
Secondary: preference types & Gains from \BSD\ are larger for participants who value cognitive ease or dislike variety. & Mechanism \(\times\) preference-type interactions. \\
Secondary: market consequences & Individual inconsistency propagates through the menus of lower-priority participants. & Market-level welfare and dispersion by mechanism (22 recoverable markets). \\
\bottomrule
\end{tabularx}
\end{table}

%may add later or add to an online appendix
%\clearpage
%\section{Experiment Material}\label{app:ExperimentMaterial}
%\includepdf[pages=-, pagecommand={\thispagestyle{plain}}]{Experiment/Screenshots.pdf}

\end{document}